\documentclass[10pt,a4paper,english,pra, twocolumn]{revtex4-1}
\usepackage[T1]{fontenc}
\usepackage[latin9]{inputenc}
\usepackage{color}
\definecolor{lyxbackgroundcolor}{rgb}{1, 1, 1}
\usepackage{array}
\usepackage{booktabs}
\usepackage{multirow}
\usepackage{amsmath}
\usepackage{amssymb}
\usepackage{graphicx}

\makeatletter

\providecommand{\tabularnewline}{\\}

\usepackage[colorlinks = true,
            linkcolor = red,
            urlcolor  = blue,
            citecolor = blue,
            anchorcolor = blue]{hyperref}
\makeatletter
\def\switch@array{}
\makeatother

\makeatother

\usepackage{babel}
\begin{document}
\title{Programmable optical parametric amplifier synthesizer for cubic phase
states and amplified Schrödinger cat states }
\author{Yusuf Turek}
\email{Corresponding author: yusufu1984@hotmail.com}

\author{Ming-Yan Sun}
\author{Xiao-Xi Yao}
\affiliation{School of Physics,Liaoning University,Shenyang,Liaoning 110036,China}
\date{\today}
\begin{abstract}
We introduce a programmable optical parametric amplifier (OPA) synthesizer
that, under a heralded photon-number-resolving framework, generates
high-fidelity cubic phase states and amplifies Schrödinger cat states.
By systematically exploring both the catalytic configuration, where
the idler input and output contain the same number of photons ($m=n$~),
and non-catalytic configurations ( $m\neq n$), we discover two qualitatively
different functionalities. First, with a coherent-state signal input,
our protocol generates cubic phase states with fidelity exceeding
$0.99$ across a broad range of ($m$, $n$) configurations. Second,
using a Schrödinger cat state as the signal input, the same framework
amplifies the cat state: an input cat with amplitude $\alpha_{in}\le1$
is transformed into an output squeezed cat with $\alpha_{out}\ge2$
while maintaining fidelity above $0.99$. The catalytic configuration
preserves the input parity and restores the idler state, whereas non-catalytic
configurations enable parity-flipping amplification with higher success
rates. Moreover, the amplified output can serve as a seed for subsequent
amplification rounds, offering a self-seeding pathway to progressively
larger cat states. Our protocol requires only moderate-gain OPA operation
and low-order photon-number-resolving detection, providing a flexible
and experimentally accessible platform for cubic phase state preparation
and amplified squeezed cat state generation.
\end{abstract}
\maketitle

\section{\label{sec:1}Introduction}

The ability to generate and manipulate non\nobreakdash-Gaussian states
of light is a cornerstone of continuous\nobreakdash-variable quantum
information processing, enabling universal quantum computation \citep{PRXQuantum.2.030204,PRXQuantum.6.010311},
fault\nobreakdash-tolerant error correction \citep{PRXQuantum.2.020101,hr5f-lvy7},
and quantum metrology beyond the standard quantum limit \citep{e27070712,https://doi.org/10.1002/qute.202100080,PhysRevA.78.063828,PhysRevA.88.013838,PhysRevA.91.013808,PhysRevA.93.033859,PhysRevLett.104.103602}.
Among the most sought\nobreakdash-after non\nobreakdash-Gaussian
resources are cubic phase states, which provide the nonlinearity required
for the cubic phase gate -- a key missing element for universal Gaussian\nobreakdash-only
architectures \citep{PhysRevA.64.012310,RevModPhys.84.621,PRXQuantum.2.010327}.

A promising approach for producing non\nobreakdash-Gaussian states
uses an optical parametric amplifier (OPA) with heralded photon detection.
Shringarpure and Franson first demonstrated that an OPA seeded with
a coherent state and a single idler photon can generate photon\nobreakdash-added
states \citep{PhysRevA.100.043802}. More recently, Erkilic et al.
showed that with vacuum idler input ( $m=0$) and heralded photon
detection, an OPA can produce cubic\nobreakdash-phase\nobreakdash-like
states with fidelities exceeding 98.5\% \citep{erkilic2025}. In these
schemes, however, the idler port is restricted to either vacuum or
a fixed Fock state. 

Furthermore, our recent work has established that this OPA heralding
architecture is more than a state generator---it is the kernel of
a reconfigurable quantum state synthesizer \citep{rkzg-sdxn,turek2026}.
By varying the idler photon numbers ($m$, $n$), the same device
implements effective high-order photon subtraction from a squeezed
vacuum, producing large-amplitude cat states with controllable parity
\citep{turek2026}. These operations constitute the synthesizer's
base modes. However, whether this platform can generate entirely new
classes of states---such as cubic phases---or amplify non-Gaussian
seeds remains an open question.

Here we fill both gaps---systematically investigating the full ($m$,
$n$) parameter space, exploring both catalytic ($m=n$) and non-catalytic
($m\neq n$) configurations. As we demonstrate below, this unified
framework yields two qualitatively new effects that are absent in
all previous OPA schemes---including our own base-mode operations
\citep{rkzg-sdxn,turek2026}. 

First, using a coherent state as the signal input, we find that high-fidelity
cubic phase states ($F\ge0.99$) can be generated across a broad range
of ($m$, $n$) configurations---not only under the catalytic condition
but also in non-catalytic settings. This result is non-trivial: coherent
states are Gaussian, while cubic phase states are highly non-Gaussian,
and no linear optical transformation can bridge the two. The OPA heralding
framework thus provides a compact, integrated nonlinear channel that
accomplishes this conversion using only photon-number-resolving detection.

Second, using an unsqueezed Schrödinger cat state as the signal input,
the same protocol amplifies the cat state: an input cat with amplitude
$\alpha_{in}\le1$ is transformed into a squeezed cat state with $\alpha_{out}\ge2$
while maintaining fidelity above $0.99$. This constitutes both amplitude
amplification and the generation of squeezing---two effects simultaneously
achieved in a single heralding step. Large-amplitude squeezed cat
states are essential resources for fault-tolerant bosonic error correction
\citep{PhysRevX.8.021054,97yt-nzg2}, yet their preparation remains
challenging. Our protocol offers a direct and experimentally accessible
route to such states starting from a small unsqueezed cat, which is
significantly easier to generate \citep{2006,PhysRevLett.101.233605}.

Notably, the catalytic configuration $m=n$ preserves the input parity
and restores the idler state, whereas non-catalytic configurations
$m\neq n$ enable parity-flipping amplification with higher success
probabilities. This flexibility allows the protocol to be tailored
for different quantum information tasks, establishing a programmable
and unified platform for non-Gaussian state engineering.

A critical question is whether the present work merely repeats the
architecture of previous studies with a different input state. We
emphasize that this is not the case. While the OPA architecture is
shared with Refs. \citep{PhysRevA.100.043802,erkilic2025,turek2026,rkzg-sdxn},
the physical effects reported here are qualitatively different. Reference
\citep{erkilic2025} restricted the idler input to vacuum ($m=0$)
and did not explore the catalysis condition $m=n$ for coherent inputs,
nor did it consider arbitrary non-catalytic configurations ($m\neq n$)
or cat state inputs at all. Thus, this work is neither a trivial extension
nor a parameter scan; it uncovers a unified and programmable OPA platform
for non-Gaussian state engineering that was previously inaccessible.

The paper is organized as follows. Section \ref{sec:2} describes
the general OPA heralding formalism for arbitrary idler input and
output photon numbers. Section \ref{sec:3} presents cubic phase state
generation from coherent state inputs under various configurations.
Section \ref{sec:4} demonstrates squeezed cat state amplification
from unsqueezed cat state inputs. In Sect. \ref{sec:5}, we discuss
effects of photon loss on system quantities including fidelity, Winger
negativity and complexity. We conclude our paper in Sect. \ref{sec:6}.
Numerical simulations of this work are done using QuTiP \citep{JOHANSSON20131234}. 

\section{\label{sec:2} The non-Gaussian state source }

\begin{figure}
\includegraphics[width=9cm]{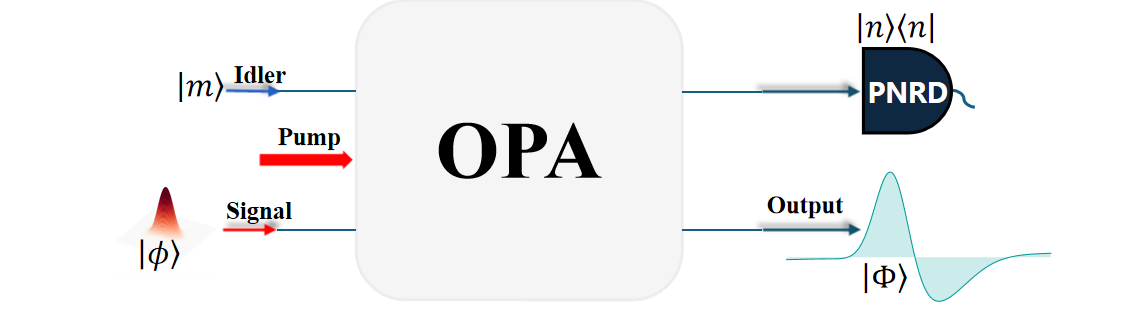}

\caption{\label{fig:1-1}Schematics of our protocol. Idler mode with Fock state
$\vert m\rangle$ and signal in state $\vert\phi\rangle$ plugged
into an optical parametric amplifier (OPA). Heralding $n$-photon
idler output gives state $\vert\Phi\rangle$ in signal output. }

\end{figure}

The schematics of our state generation protocol is showed in Fig.
\ref{fig:1-1}. Our protocol employs an optical parametric amplifier
(OPA) with two input ports: the signal port, which carries the input
state $\vert\psi\rangle$ (either a coherent state or a Schrödinger
cat state in this present work), and the idler port, which is prepared
in a Fock state $\vert m\rangle$ with a definite photon number $m$.
The OPA interaction, characterized by a tunable gain parameter $g$,
entangles the two modes and induces photon transfers between them.
At the idler output, we perform photon\nobreakdash-number resolving
detection (PNRD) and postselect events in which exactly $n$ photons
are detected, where $n$ can be any non\nobreakdash-negative integer,
not necessarily equal to $m$. This conditional measurement projects
the signal mode into a non\nobreakdash-Gaussian state. 

The action of the OPA can be described by the two\nobreakdash-mode
squeezing operator 
\begin{equation}
S(\tau)=\exp\left(\tau^{\ast}ab-\tau a^{\dagger}b^{\dagger}\right),\label{eq:e}
\end{equation}
 where $\tau=\varrho e^{i\delta}$ and $g=\cosh\varrho$ denote the
non-linear and gain parameters of the OPA, $G=\tanh\tau=\sqrt{g^{2}-1}/g$,
and $a$ ($a^{\dagger}$) $b$ ($b^{\dagger}$) are the annihilation
(creation) operators of signal and idler modes, respectively. The
normal ordered decomposition of $S(\tau)$ can be written as 
\begin{equation}
S(\tau)=\frac{1}{g}e^{-Ga^{\dagger}b^{\dagger}}g^{-(a^{\dagger}a+b^{\dagger}b)}e^{Gab}.\label{eq:2-1}
\end{equation}
 The initial joint state is chosen as $\vert\phi\rangle\otimes\vert m\rangle$,
with an arbitrary signal state $\vert\phi\rangle$ and a Fock state
$\vert m\rangle$ in the idler. After the OPA interaction, a photon\nobreakdash-number\nobreakdash-resolving
measurement on the idler mode postselects the outcome $n$ , which
projects the signal mode into the unnormalised state $\vert\psi\rangle_{m,n}=\langle n\vert S(\tau)\vert\phi\rangle\vert m\rangle$.
To evaluate this expression, we apply the three factors of the decomposition
successively to the initial state and keep only the terms that lead
to exactly $n$ photons in the idler. resulting expression takes the
compact form 

\begin{equation}
\vert\Psi\rangle_{\psi,m,n}=\sum^{m}_{k=0}H(k,m,n)(a^{\dagger})^{n-m+k}g^{-a\dagger a}a^{k}\vert\psi\rangle\label{eq:1}
\end{equation}
 where 
\begin{equation}
H(k,m,n)=\frac{G^{^{n-m+2k}}}{g^{m-k+1}}\frac{(-1)^{n-m+k}\sqrt{n!m!}}{k!(n-m+k)!(m-k)!}.\label{eq:2}
\end{equation}
 The power of $G$ is obtained from the product $G^{k}.(-G)^{n-m+k}$,
and the denominator contains the factorials from the expansions as
well as an extra power of $g$ that originates from the idler attenuation
and the initial $1/g$. 

The heralded signal state in Eq. (\ref{eq:1}) is a coherent superposition
over $k=0,...,m$ of terms in which the input state $\vert\phi\rangle$
is sequentially processed by three operations: first, $k$ photons
are subtracted via $a^{k}$; second, the resulting state undergoes
noiseless attenuation through $g^{-a^{\dagger}a}$ ; and third, $n-m+k$
photons are added via $(a^{\dagger})^{n-m+k}$. This right\nobreakdash-to\nobreakdash-left
ordering is essential because the three operators do not commute,
and it arises naturally from the normal\nobreakdash-ordered decomposition
of the two\nobreakdash-mode squeezing unitary. Thus, the protocol
implements a programmable sequence of photon subtraction, amplitude
scaling, and photon addition, with the net photon\nobreakdash-number
change determined by the difference between the detected and input
idler photon numbers.

The resulting heralded signal state $\vert\Psi\rangle_{\psi,m,n}$
depends on the chosen parameters ($m$, $n$, $g$) and the input
state. By scanning these parameters, we can tailor the output to match
various target non\nobreakdash-Gaussian states. In this work, we
demonstrate two prominent applications: for a coherent input, appropriate
choices of ($m$, $n$, $g$) yield cubic phase states with fidelity
exceeding 0.99, while for a cat\nobreakdash-state input, the protocol
transforms a small\nobreakdash-amplitude unsqueezed cat into a larger\nobreakdash-amplitude
squeezed cat with extreme high fidelity. Notably, the catalytic case
$m=n$ is a special subclass that preserves the input parity and restores
the idler state, but it is not the only working configuration; non\nobreakdash-catalytic
settings can also offer higher success probabilities. Thus, our scheme
provides a general and programmable non\nobreakdash-Gaussian state
synthesizer, with arbitrary $m$, $n$ offering flexible trade\nobreakdash-offs
between fidelity, success probability, and experimental resources. 

The norm square of this unnormalised state gives the success probability
$P_{m,n}=Tr\left(\vert\Psi\rangle_{\psi,m,n}\langle\Psi\vert\right)$
for the heralding event, which is the conditional probability that
the idler output contains exactly $n$ photons given that the idler
input was prepared in $\vert m\rangle$ and the signal input was $\vert\psi\rangle$.
However, in a realistic experimental implementation, the total probability
per trial for successfully generating the heralded state must also
account for the probability of preparing the initial $m$-photon Fock
state at the idler input. If this Fock state is generated by a separate
spontaneous parametric down-conversion (SPDC) source with gain parameter
$g^{\prime}$, then the preparation probability is $P_{Fock}(m,g^{\prime})=\vert_{s}\langle m\vert(\mathbb{I}\otimes\vert m\rangle\langle m\vert)S(\tau^{\prime})(\vert0\rangle_{a}\otimes\vert0\rangle_{b})\vert^{2}.$
The overall success probability per trial is therefore the product
of this Fock-state preparation probability, the conditional OPA heralding
probability $P_{m,n}$, and the detection efficiency $\eta_{det}$
of the photon-number-resolving detector: $P_{trial}=\eta_{det}P_{Fock}(m,g^{\prime})P_{m,n}$.
This factorization holds because the two heralding stages---Fock-state
generation and OPA interaction---are independent when the idler mode
is losslessly routed between them. So far, PNRDs with high quantum
efficiency ( $\eta_{det}\ge0.95$) have been reported \citep{Lita:08,Fukuda:11,2012,Zhang:24}.
Hence, in practice, the dominant contribution to low $P_{trial}$
comes from either the Fock-state preparation (especially for large
$m$ where multi-photon SPDC is inefficient) or the OPA heralding
probability (which decreases with increasing detected photon number
$n$). Despite these limitations, the single-trial probabilities we
obtain are comparable to or even higher than those reported in other
heralded non-Gaussian state generation schemes, and the low values
for high-$m$ configurations can be overcome by operating the pump
laser at high repetition rates \citep{Jornod:23,Wakui:20,Zhang:24}.
In the following sections, we will present explicit results of $P_{trial}$
for the specific configurations that generate cubic phase states and
amplified cat states, demonstrating that our protocol is experimentally
feasible with current technology. 

Before analyzing specific input states, we recall the definitions
of two important families of states that will be used throughout this
work. A coherent state is obtained by applying the displacement operator
$D(\alpha)=\exp\left(\alpha a^{\dagger}-\alpha^{\ast}a\right)$ to
the vacuum: 
\begin{align}
\vert\alpha\rangle & =D(\alpha)\vert0\rangle=e^{-\vert\alpha\vert^{2}/2}\sum^{\infty}_{p=0}\frac{\alpha^{p}}{\sqrt{p!}}\vert0\rangle.\label{eq:coh}
\end{align}
 A cat state is a superposition of two coherent states with opposite
amplitudes and a relative phase: 

\begin{equation}
\vert Cat\rangle_{\alpha,\theta}=N^{-1/2}_{\theta}\left(\vert\alpha\rangle+e^{i\theta}\vert-\alpha\rangle\right),\label{eq:cat}
\end{equation}
 where $N_{\theta}=2\left(1+e^{-2\alpha^{2}}\cos\theta\right)$ ensures
normalisation, and the even and odd cat states (ECS,OCS) correspond
to $\theta=0$ and $\theta=\pi$, respectively. 

As mentioned above the properties of output signal state $\vert\Psi\rangle_{\psi,m,n}$
depends on the OPA gain $g$, the input and detected photon numbers
$m$, $n$, and the input signal state $\vert\psi\rangle$. For general
parameters the state is a complex superposition of many terms, but
in certain limiting cases the expression simplifies considerably and
admits a clear physical interpretation.

(i) When $m=n=0$, the idler is initially in the vacuum and we postselect
on zero photons at the output; the sum in Eq. (\ref{eq:1}) reduces
to the single term $k=0$ yielding $\vert\Psi\rangle_{\psi,0,0}\propto g^{-a^{\dagger}a}\vert\psi\rangle$.
As in investigated in Ref. \citep{PhysRevA.96.042307} this operation
is known as noiseless attenuation. It scales the amplitude of every
Fock component without changing the photon\nobreakdash-number parity
or introducing additional photons. For a coherent state input, $g^{-a^{\dagger}a}\vert\alpha\rangle\propto\vert\alpha/g\rangle$,
so the output remains a coherent state with its amplitude reduced
by a factor of $g$. For a cat state input, the same attenuation acts
on each coherent component individually, producing $\propto\vert Cat\rangle_{\alpha/g,\theta}$
with the same relative phase $\theta$ but a smaller amplitude.

(ii) For $m=0$ and $n\neq0$, i.e., vacuum idler input but heralding
on $n$ photons, only the $k=0$ term survives and the output becomes
$\vert\Psi\rangle_{\psi,0,n}\propto(a^{\dagger})^{n}g^{-a^{\dagger}a}\vert\psi\rangle.$
Thus the protocol effectively adds $n$ photons to the input state
after an attenuation, which is precisely the photon\nobreakdash-added
attenuated state studied in previous works \citep{PhysRevA.78.063811,Park:16,2024,2025K}.

(iii) In the complementary case $m\neq0$ and $n=0$, the detected
photon number is zero while the idler initially contains $m$ photons.
The condition $n-m+k$ forces $k=m$, so only the last term of the
sum contributes, giving $\vert\Psi\rangle_{\psi,m,0}\propto g^{-a^{\dagger}a}a^{m}\vert\psi\rangle$.
This corresponds to $m$-photon subtraction followed by noiseless
attenuation, a process that can generate non\nobreakdash-Gaussian
features such as Wigner negativity when applied to a suitable input
state \citep{2006,PhysRevLett.101.233605,PhysRevA.82.031802}.

These three special cases already illustrate the versatility of the
OPA heralding scheme: by choosing ($m$, $n$) we can implement noiseless
attenuation, photon addition, or photon subtraction within a single
unified framework. In the following sections, we will discuss in detail
the cases where the signal input is a coherent state and a Schrödinger
cat state, respectively, and demonstrate how the general heralded
state $\vert\Psi\rangle_{\psi,m,n}$ can be engineered to produce
high\nobreakdash-fidelity cubic phase states and amplified squeezed
cat states through appropriate choices of the parameters $m$, $n$
and the OPA gain $g$. 

\section{\label{sec:3}Cubic Phase State Generation from Coherent state input}

\begin{figure*}
\includegraphics[width=16cm]{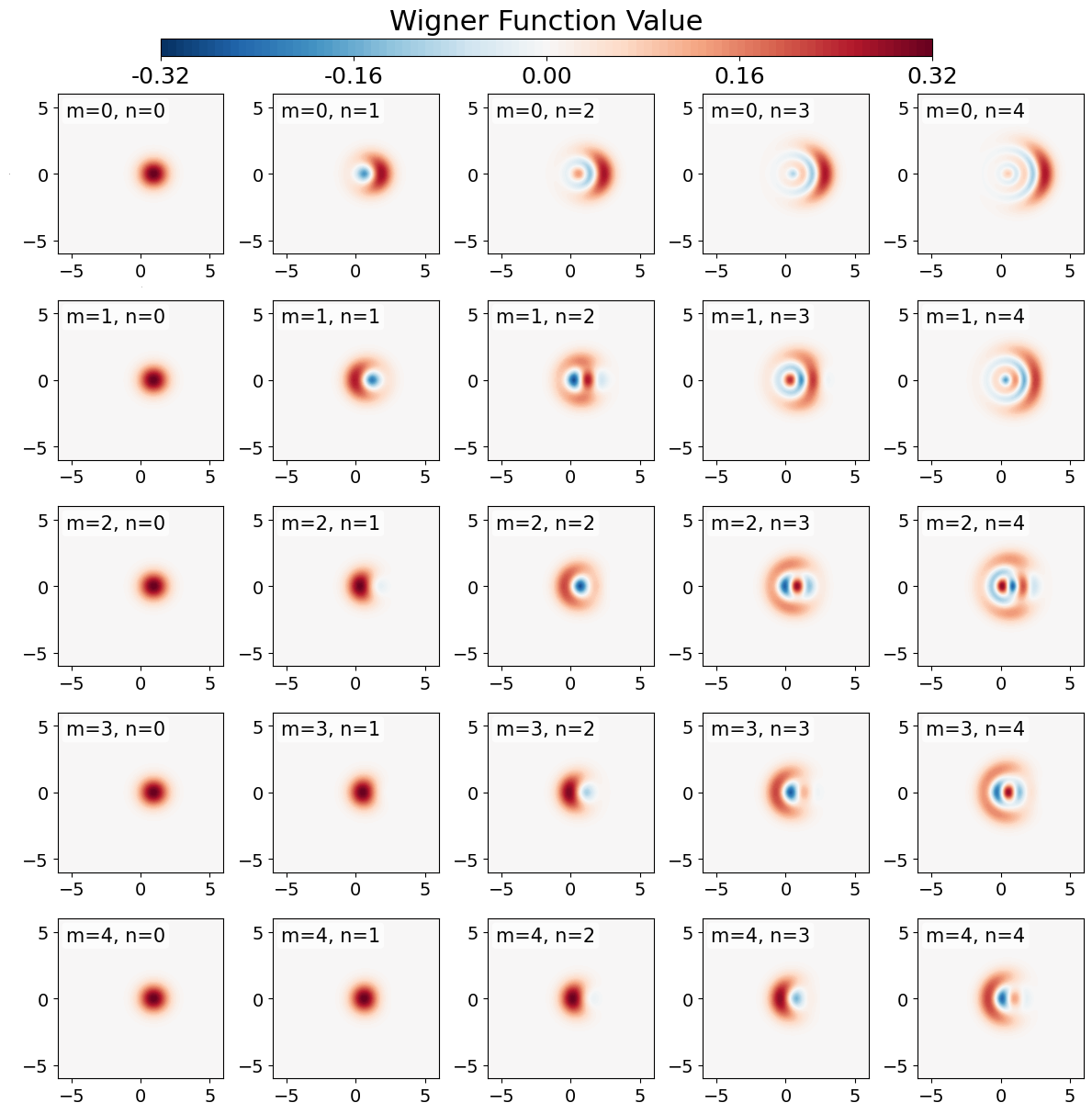}

\caption{\label{fig:1}Wigner functions of the heralded signal $\vert\Psi^{\prime}\rangle_{coh,n,n}$
for coherent state input with various configurations ($m$, $n$),
with fixed $\alpha=1$ and $g=1.5$. The emergence of negative regions
indicates non\protect\nobreakdash-classicality. }
\end{figure*}

We first consider a coherent state input $\vert\psi\rangle=\vert\alpha\rangle=D(\alpha)\vert0\rangle$.
Substituting into Eq. (\ref{eq:1}) using the identity $g^{-a^{\dagger}a}\vert\alpha\rangle\propto\vert\alpha/g\rangle$,
we obtain the unnormalized heralded state 
\begin{align}
\vert\Psi\rangle_{coh,m,n} & =\exp\left[-\frac{G^{2}\vert\alpha\vert^{2}}{2}\right]\sum^{m}_{k=0}H(k,m,n)\alpha^{k}(a^{\dagger})^{n-m+k}\vert\frac{\alpha}{g}\rangle,\label{eq:3}
\end{align}
where we have used $a^{k}\vert\alpha/g\rangle=\left(\alpha/g\right)^{k}\vert\alpha/g\rangle$,
valid for a coherent state. Expanding the photon\nobreakdash-addition
operator $\left(a^{\dagger}\right)^{n-m+k}$ in the Fock basis and
applying the displacement operator $D(\alpha/g)$, the state can be
rewritten in the compact form
\begin{align}
\vert\Psi\rangle_{coh,m,n} & =D(\frac{\alpha}{g})\sum^{m}_{k=0}\sum^{n-m+k}_{l=0}C(m,n,k,l)\vert l\rangle,\label{eq:co2}
\end{align}
with the coefficients 
\begin{align}
C(m,n,k,l) & =\exp\left[-\frac{G^{2}\vert\alpha\vert^{2}}{2}\right]\times\nonumber \\
 & H(k,m,n)\alpha^{k}\frac{(n-m+k)!(\frac{\alpha^{\ast}}{g})^{n-m+k-l}}{\sqrt{l!}(n-m+k-l)!}.
\end{align}
 This representation shows that the heralded state is a globally displaced
superposition of Fock states, with the displacement amplitude reduced
by the OPA gain factor $g$. The Wigner functions of $\vert\Psi\rangle_{coh,m,n}$
for various ($m$, $n$) configurations with fixed $g=1.5$ are plotted
in Fig. \ref{fig:1}, where the emergence of negative regions---signatures
of non\nobreakdash-classicality---is clearly visible for configurations
with non\nobreakdash-zero photon addition.

When the model parameters are set to the symmetric case $m=n$, the
general heralded state simplifies considerably and acquires a clear
physical interpretation. The resulting unnormalized output state can
be written as

\begin{align}
\vert\Psi^{\prime}\rangle_{coh,n,n} & =D\left(\frac{\alpha}{g}\right)\sum^{n}_{l=0}\mathcal{C}_{l}\vert l\rangle,\label{eq:8}
\end{align}
 where the coefficients $\mathcal{C}_{l}=\sum^{n}_{k=l}C_{n,k,l}$
are given by 
\begin{equation}
C_{n,k,l}=\exp\left[-\frac{G^{2}\vert\alpha\vert^{2}}{2}\right]\frac{1}{g^{n-l+1}}\frac{(-1)^{k}n!G^{2k}}{k!(n-k)!}\frac{\alpha^{k}(\alpha^{\ast})^{k-l}}{\sqrt{l!}(k-l)!}.\label{eq:c1}
\end{equation}

The output state in Eq. (\ref{eq:8}) represents a displaced, photon-number-truncated
non-Gaussian state. Its structure --- a coherent superposition of
Fock states up to $\vert n\rangle$ globally displaced by $\alpha/g$
--- is tunable via the scaling parameter $g$. Unlike standard photon-added
coherent states \citep{2024}, this state does not amplify the input
coherent state for $g>1$. Instead, it provides a flexible platform
for generating non-Gaussian states with controllable photon-number
distributions and Wigner-function negativity. 

Among non-Gaussian resources, cubic phase states play a particularly
important role in continuous-variable quantum information processing
\citep{PhysRevA.64.012310,PhysRevLett.82.1784}. A cubic phase state
(CPS) is defined as
\begin{equation}
\vert\gamma^{\prime}\rangle=e^{i\gamma^{\prime}x^{3}}\vert0\rangle_{p},
\end{equation}
 where $\gamma^{\prime}\in\mathbb{R}$ is the strength of the cubic
nonlinearity (hereafter referred to as the cubicity), $x=(a+a^{\dagger})/\sqrt{2}$
is the position quadrature, and $\vert0\rangle_{p}$ is the zero-momentum
eigenstates. Since the ideal zero-momentum eigenstate is unphysical,
experimentally realizable cubic phase states are typically prepared
with finite squeezing:
\begin{equation}
\vert\Psi_{target}\rangle=e^{i\gamma^{\prime}x^{3}}S(r)\vert0\rangle,\label{eq:5}
\end{equation}
 where $S(r)=\exp\left[\frac{r}{2}(a^{2}-a^{\dagger2})\right]$ the
squeezing operator and $r\in\mathbb{R}$ is the squeezing parameter.
All squeezing values $r$ used in this work can be converted to the
decibel (dB) scale using the formula $r_{dB}=\frac{20}{\ln10}r\approx8.686r.$
CPSs are essential resources for universal continuous-variable quantum
computation, as they enable the implementation of the cubic phase
gate $e^{i\gamma^{\prime}x^{3}}$ which, together with Gaussian operations,
forms a universal gate set \citep{PhysRevLett.82.1784}. Despite their
importance, generating high-fidelity cubic phase states has remained
a long-standing challenge. 

So far several approaches have been proposed to effectively generate
CPSs. Arzani et al. \citep{PhysRevA.95.052352} proposed cascaded
measurement protocols to cubic phase states using single-photon detection,
but the success probability decreases rapidly with the number of required
steps. Kerr nonlinearity has been explored as a route to the cubic
phase gate \citep{PhysRevA.88.053816,PhysRevLett.124.240503}, but
its weak strength at the single-photon level has motivated alternative
approaches using Gaussian enhancement \citep{PhysRevLett.124.240503}
or conditional state preparation \citep{PhysRevA.88.053816}. Gaussian
conversion protocols transform a pre-existing non-Gaussian state---the
trisqueezed state---into a cubic phase state using only squeezing
and displacement, achieving fidelities up to $F=0.997$, but this
approach requires a complex non-Gaussian seed generated by challenging
three-photon down-conversion techniques \citep{PRXQuantum.2.010327}.
Interferometer-based heralded schemes have been explored using a Fock
state $\vert2\rangle$ and a coherent state as inputs, followed by
projection onto $\vert2\rangle$. Kashyap et al. \citep{Kashyap_2026}
demonstrated that such a scheme can generate cubic phase states with
fidelity $F\sim0.99$ , but at an extremely low success probability
$P\sim10^{-7}$ when high fidelity is required. Balancing the beamsplitter
to improve the success probability reduces the fidelity to $F\sim0.97$. 

Most recently, Erkilic et al. \citep{2025K} demonstrated a significant
advance to generate CPS using an OPA with heralded photon detection.
In their scheme, the idler port is seeded with vacuum ($m=0$), while
a coherent state is injected into the signal port. By performing PNRD
on the idler output and heralding on $n$ photons ($n\ge1$), they
showed that cubic-phase-like states can be generated with fidelities
exceeding $98.5\%$. This work is notable because it generates non-Gaussian
states using only Gaussian inputs (coherent states) and a single OPA,
without requiring pre-existing non-Gaussian resources such as Fock
states or trisqueezed states. However, the idler port in Erkilic et
al.'s scheme is restricted to vacuum input ($m=0$). The more general
case where the idler is seeded with a Fock state ($m\ge1$) and in
particular the catalysis configuration where the idler input and output
contain the same number of photons ($m=n$) was not explored. 

In this work, we generalize the OPA heralding framework to arbitrary
idler input Fock states ($m\ge0$) and arbitrary heralded photon numbers
($n\ge0$). We systematically investigate the full parameter space
of ($m$,$n$) configurations and discover that cubic phase states
can be generated across a wide range of configurations --- not only
under the catalysis condition $m=n$, but also for $m\neq n$ cases. 

\begin{table*}
\caption{\label{tab:1} Optimized parameters ( $g$, $\gamma$, $r$, $\alpha$,
$r_{corr}$, $\beta_{corr}$ , fidelity $F$, and success probability
$P_{trial}$ ) for cubic phase state (CPS) generation from coherent
state input for various ($m$, $n$) configurations.}

\begin{tabular}{>{\centering}p{1cm}>{\centering}p{1cm}>{\centering}p{1cm}>{\centering}p{1cm}>{\centering}p{1cm}>{\centering}p{1.5cm}>{\centering}p{1cm}>{\centering}p{1.5cm}>{\centering}p{1.2cm}>{\centering}p{2.2cm}}
\toprule 
\multicolumn{2}{c}{$(m,n)$} & $g$ & $\gamma^{\prime}$ & $r$ & $\alpha$ & $r_{corr}$ & $\beta_{corr}$ & $F$ & $P_{total}$\tabularnewline
\midrule
\multicolumn{2}{c}{$(0,4)$} & $1.46$ & $0.103$ & $0$ & $-2.145i$ & $0.521$ & $2.963i$ & $0.997$ & $8.28\times10^{-2}$\tabularnewline
\multicolumn{2}{c}{$(1,1)$} & $1.08$ & $0.100$ & $0$ & $-4.136i$ & $0.203$ & $4.406i$ & $0.975$ & $1.17\times10^{-2}$\tabularnewline
\multicolumn{2}{c}{$(1,4)$} & $1.15$ & $0.146$ & $-0.01$ & $-1.448i$ & $0.589$ & $2.493i$ & $0.989$ & $8.82\times10^{-3}$\tabularnewline
\multicolumn{2}{c}{$(1,5)$} & $1.07$ & $0.100$ & $0$ & $-2.260i$ & $0.469$ & $3.401i$ & $0.991$ & $3.75\times10^{-4}$\tabularnewline
\multicolumn{2}{c}{$(2,2)$} & $2.06$ & $0.100$ & $0$ & $-3.676i$ & $0.399$ & $2.786i$ & $0.993$ & $1.13\times10^{-5}$\tabularnewline
\multicolumn{2}{c}{$(2,5)$} & $1.05$ & $0.100$ & $0$ & $-1.614i$ & $0.504$ & $2.703i$ & $0.995$ & $3.80\times10^{-4}$\tabularnewline
\multicolumn{2}{c}{$(3,3)$} & $2.34$ & $0.128$ & $0$ & $-3.933i$ & $0.527$ & $3.020i$ & $0.992$ & $6.69\times10^{-7}$\tabularnewline
\multicolumn{2}{c}{$(4,4)$} & $2.47$ & $0.166$ & $-0.01$ & $-3.960i$ & $0.647$ & $3.258i$ & $0.987$ & $1.19\times10^{-7}$\tabularnewline
\bottomrule
\end{tabular}
\end{table*}

The optimisation procedure for generating cubic phase states proceeds
as follows. For a given ($m$, $n$) configuration and input coherent
amplitude $\alpha$, we compute the heralded state $\vert\Psi\rangle_{coh,m,n}$
from Eq. ( \ref{eq:3}) and then apply a squeezing operation $S(r_{corr})=\exp[\frac{r_{corr}}{2}(a^{2}-a^{\dagger2})]$
followed by a displacement $D(\beta_{corr})$ to match the target
cubic phase state $\vert\Psi_{target}\rangle$. The fidelity $F=\vert\langle\Psi_{target}\vert S(r_{corr})D(\beta_{corr})\vert\Psi\rangle_{coh,m,n}\vert^{2}$
is maximised over the parameters ( $g$, $\alpha$, $r$, $\gamma^{\prime}$,
$r_{corr}$, $\beta_{corr}$ ), where the OPA gain $g$ and input
amplitude $\alpha$ determine the heralded state, while $r_{corr}$
and $\beta_{corr}$ are the Gaussian correction parameters, and $\gamma^{\prime}$
is the target cubicity. The numerical optimisation is performed using
gradient\nobreakdash-based methods implemented in QuTiP \citep{JOHANSSON20131234}.

Table \ref{tab:1} presents the optimised parameters and resulting
fidelities for several representative configurations. We first focus
on the catalytic cases $m=n$, which are the central new contribution
of this work. The configurations ($1,$ $1$), ($2$, $2$), ($3$,
$3$), and ($4,$$4$) all yield fidelities close to or exceeding
$0.99$ with appropriately chosen OPA gains and input amplitudes.
For comparison, we also include the ($0,$$4$) configuration which
was previously studied by Erkilic et al. \citep{2025K} and is included
here as a reference; it achieves a fidelity of $0.997$ with a significantly
higher success probability of $8.28\times10^{-2}$. In addition to
the catalytic and vacuum-input cases, we have identified several non-catalytic
configurations with $m\neq n$ and $m\neq0$ that also produce high-fidelity
cubic phase states, such as ($1$, $4$), ($1$, $5$), and ($2$,
$5$), with fidelities of $0.989$, $0.991$, and $0.995$, respectively.
These results demonstrate that high-fidelity cubic phase states can
be generated across a broad range of ($m$, $n$) configurations,
not only under the catalytic condition but also in non-catalytic settings,
offering flexibility in balancing fidelity and success probability.
In~ our protocol, the target squeezing parameter $r$ (fourth column)
is near zero (approximately $-0.01$to $0$) for all cases. This indicates
that the our optimization process forced to take it close zero values
for giving good approximation to the target state for different configurations. 

\begin{figure}
\includegraphics[width=8.6cm]{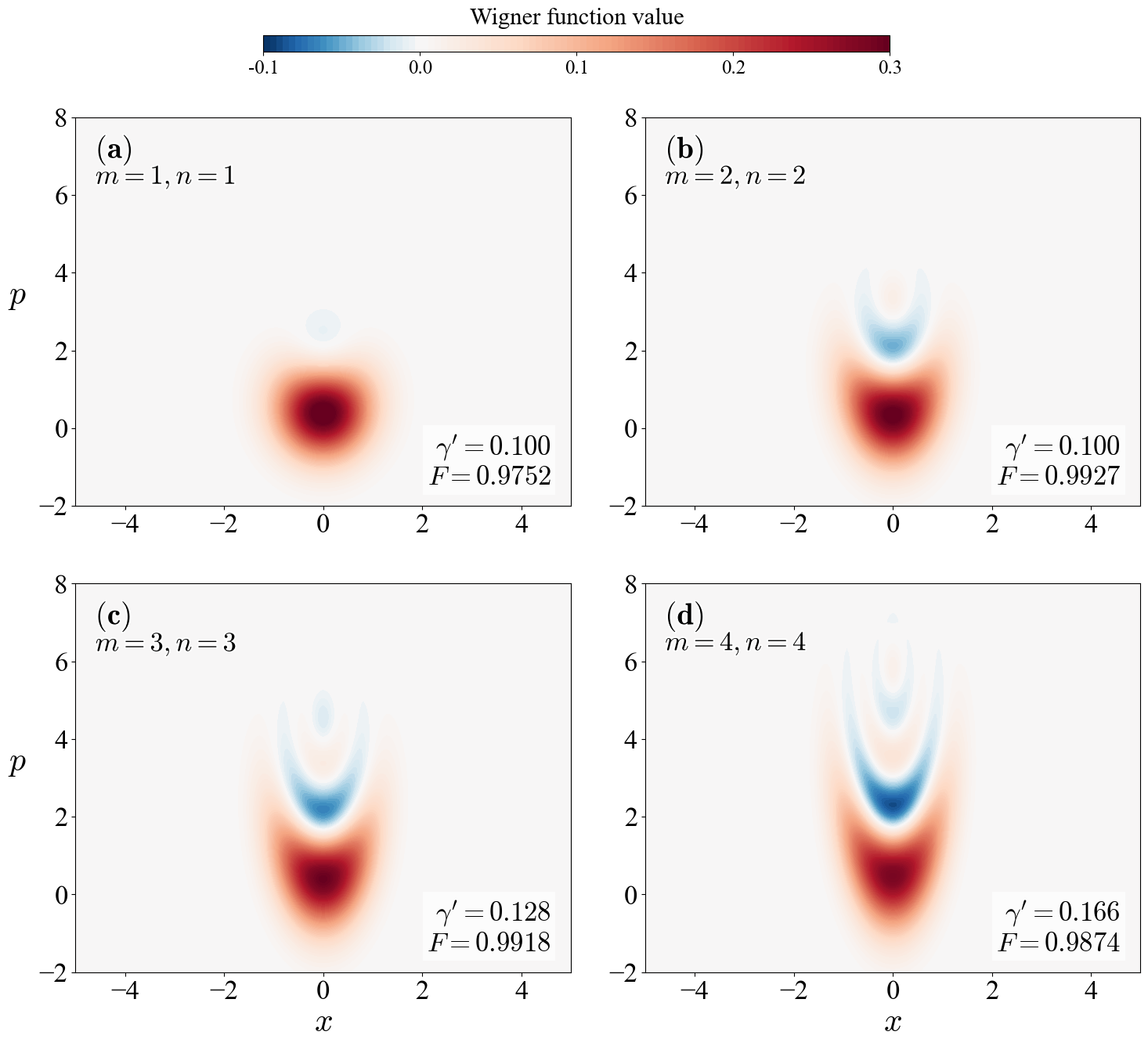}

\caption{\label{fig:3}Wigner functions of the optimized output states for
catalytic configurations $m=n=1,2,3,4$ with coherent input, showing
cubic phase state (CPS) features.}
\end{figure}

\begin{figure}
\includegraphics[width=8.6cm]{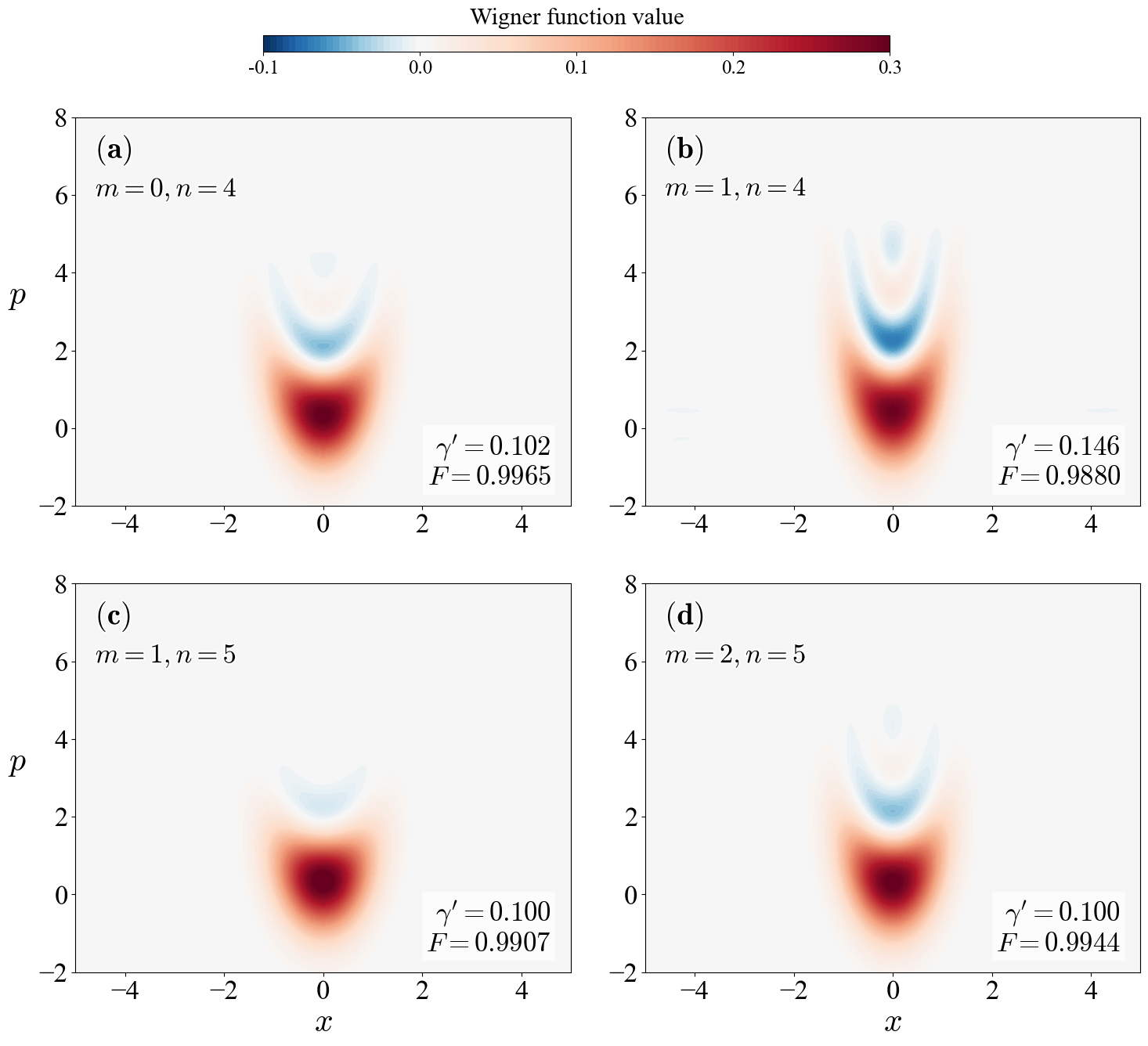}

\caption{\label{fig:4} Wigner functions for non\protect\nobreakdash-catalytic
configurations ($1,$$4$), ($1,$$5$), ($2$, $5$) and the reference
case ($0$, $4$) with coherent input, optimized to generate high\protect\nobreakdash-fidelity
cubic phase states (CPSs). }
\end{figure}

The Wigner functions of the optimized heralded output states for the
catalytic cases $m=n=1,2,3,4$ are shown in Fig. \ref{fig:3}, where
the characteristic non\nobreakdash-Gaussian features---negative
regions and asymmetric distortions---are clearly visible and closely
match those of the ideal CPSs. For comparison, Fig. \ref{fig:4} shows
the Wigner functions for several non\nobreakdash-catalytic configurations,
including ($1$, $4$), ($1$, $5$), and ($2$, $5$), and the reference
case ($0$, $4$), illustrating the rich variety of non\nobreakdash-Gaussian
states accessible within our framework. 

A key observation from Table \ref{tab:1} is that the optimal OPA
gain $g$ increases with the catalysis order. This trend reflects
the fact that higher\nobreakdash-order catalysis requires stronger
nonlinear interactions to generate the higher\nobreakdash-Fock components
necessary for the cubic phase state. The optimal input coherent amplitude
$\alpha$ (fifth column) is purely imaginary for all configurations.
Since the imaginary part of $\alpha$ determines the displacement
along the $p$-quadrature (momentum), a negative imaginary amplitude
shifts the Wigner function toward the negative $p$ direction. This
initial displacement, combined with subsequent Gaussian corrections,
allows the state to be positioned in phase space to match the target
cubic phase state. 

The Gaussian correction parameters---the real squeezing parameter
$r_{corr}$ (sixth column) and the complex displacement $\beta_{corr}$(seventh
column) take values where $r_{corr}$ is positive and $\beta_{corr}$
is purely imaginary for all configurations. The positive imaginary
part of $\beta_{corr}$ shifts the Wigner function in the positive
$p$ compensating for the negative $p$-displacement introduced by
the initial $\alpha$. Notably, $r_{corr}$ increases with the catalysis
order, from $0.203$ for ($1$,$1$) to $0.647$ for ($4$, $4$),
indicating that higher\nobreakdash-order catalysis produces states
that require stronger squeezing correction to match the target. We
note that implementing the corrective squeezing on a displaced non-Gaussian
state is experimentally demanding; nevertheless, it has been demonstrated
in Ref. \citep{PhysRevLett.113.013601} and is becoming increasingly
feasible with modern homodyne feedback techniques. The Wigner functions
in Figs. \ref{fig:3} and \ref{fig:4} all exhibit reflection symmetry
about the $x$-axis, i.e., $W(x,p)=W(x,-p)$. This symmetry follows
directly from the fact that both the input displacement $\alpha$
and the correction displacement $\beta_{corr}$ are purely imaginary,
so all displacements occur exclusively along the $p$-quadrature,
leaving the $x$-axis symmetry intact. Importantly, this symmetry
matches that of the ideal CPS, confirming that our phase-space alignment
strategy successfully reproduces the correct symmetry structure of
the target state.

The success probabilities $P_{trial}$ (last column) vary widely across
configurations. The reference ($0$, $4$) case offers the highest
value, on the order of $10^{-2}$. For catalytic configurations range,
$P_{trial}$ ranges from $10^{-2}$ for ($1$, $1$) down to $10^{-7}$
for ($4$, $4$), reflecting the increasing experimental difficulty
of preparing higher\nobreakdash-order Fock states and detecting higher
photon numbers. Non\nobreakdash-catalytic configurations such as
($1$, $4$), ($1$, $5$), and ($2$, $5$) provide intermediate
probabilities on the order of $10^{-3}$ to $10^{-4}$, offering useful
trade\nobreakdash-offs between fidelity and generation rate. As discussed
in Sec. \ref{sec:2}, these low probabilities can be overcome by operating
at high repetition rates, making all configurations experimentally
feasible.

In summary, our OPA heralding framework provides a versatile platform
for generating high\nobreakdash-fidelity ( $>0.99$) CPSs from coherent
light across a broad range of ($m$, $n$) configurations. Compared
with the deterministic Gaussian conversion protocol of Zheng et al.
\citep{PRXQuantum.2.010327}, which requires a complex trisqueezed
state as a non\nobreakdash-Gaussian seed, our scheme uses only low\nobreakdash-order
Fock states (typically $\vert1\rangle$ or $\vert2\rangle$), which
are significantly easier to prepare, while achieving comparable or
higher fidelities. Compared with the vacuum\nobreakdash-input OPA
scheme of Erkilic et al. \citep{2025K} ($m=0$), which is restricted
to vacuum idler input, our work generalises the framework to arbitrary
Fock\nobreakdash-state inputs and identifies the catalytic condition
$m=n$ as a new route to high\nobreakdash-fidelity CPSs with the
added advantages of parity preservation and idler restoration. The
catalytic configurations offer these unique features at the cost of
lower success probabilities for higher orders, while non\nobreakdash-catalytic
configurations provide intermediate trade\nobreakdash-offs. This
flexibility allows experimentalists to choose the optimal configuration
balancing fidelity, success probability, and resource requirements
for their specific tasks.

\section{\label{sec:4}Amplified cat states}

We now turn to the second application of our OPA heralding framework:
the amplification of Schrödinger cat states. Specifically, we consider
an input cat state $\vert Cat\rangle_{\alpha,\theta}$ in the signal
port, with $\theta=0$ for an ECS and $\theta=\pi$ for an OCS, and
investigate how the heralding process transforms it under various
($m$, $n$) configurations.

Substituting the cat state input into the general heralded state of
Eq. (\ref{eq:1}) and using the identity $e^{-a^{\dagger}a}\vert Cat\rangle_{\alpha,\theta}\propto\vert Cat\rangle_{\alpha/g,\theta}$,
we obtain the unnormalised output state 
\begin{equation}
\vert\Psi\rangle_{cat,m,n}=\sum^{m}_{k=0}C^{\prime}_{k}\vert Cat,n-m+k\rangle_{\alpha/g,\theta+\pi k},\label{eq:14}
\end{equation}
where we have used $a^{k}\vert Cat\rangle_{\alpha/g,\theta}=(\alpha/g)^{k}\vert Cat\rangle_{\alpha/g,\theta+\pi k}$,
\[
C^{\prime}_{k}=\exp\left[-\frac{G^{2}\vert\alpha\vert^{2}}{2}\right]\mathcal{N}^{-1}_{n-m+k}N^{-1/2}_{\theta}H(k,m,n)\alpha^{k}
\]
and 
\begin{equation}
\vert Cat,l\rangle_{\alpha,\theta}=\mathcal{N}_{l}(a^{\dagger})^{l}\left(\vert\alpha\rangle+e^{i\theta}\vert-\alpha\rangle\right)\label{eq:15}
\end{equation}
is the $l$-photon-added cat state with normalization 
\begin{equation}
\mathcal{N}_{l}=2\left(l!\right)\left[L_{l}\left(-\vert\alpha\vert^{2}\right)+L_{l}\left(\vert\alpha\vert^{2}\right)e^{-2\vert\alpha\vert^{2}}\cos\left(\theta+\pi l\right)\right]^{-1/2},\label{eq:16}
\end{equation}
with $L_{l}(x)$ is the Laguerre polynomial of order $l$. This representation
reveals that the heralded state is a superposition of $m+1$ terms,
each corresponding to the input cat state with a different number
of added photons and a phase shift $\pi k$ acquired by the relative
phase between the two coherent components.

A key property of this transformation is the parity selection rule.
The parity of the heralded output state is determined by the parity
of the input cat state multiplied by the factor $(-1)^{m+n}$. Specifically,
for an ECS input ($\theta=0$), the output parity is $(-1)^{m+n}$;
for an OCS input ($\theta=\pi$), the output parity is $(-1)^{m+n+1}$. 

When the catalytic condition $m=n$ is imposed, the expression simplifies
considerably:
\begin{equation}
\vert\Psi^{\prime}\rangle_{cat,n,n}=\sum^{n}_{k=0}C^{\prime\prime}_{k}\vert Cat,k\rangle_{\alpha/g,\theta+\pi k},\label{eq:12}
\end{equation}
where $C^{\prime\prime}_{k}=\exp\left[-\frac{G^{2}\vert\alpha\vert^{2}}{2}\right]\mathcal{N}^{-1}_{k}H(k,n,n)N^{-1/2}_{\theta}\alpha^{k}$.
Under this condition, the parity of the input cat state is strictly
preserved: an ECS input yields an ECS output, and an OCS input yields
an OCS output. This parity preservation is a distinct feature of our
catalysis protocol. In previous cat-state amplification schemes, such
as the homodyne-based breeding protocol of Sychev et al. \citep{RN87},
an OCS is converted into an ECS---i.e., the parity is flipped. By
contrast, our OPA catalysis scheme preserves the parity of the input
cat state, offering a new degree of control in cat-state engineering.

In this work, we primarily investigate the similarity between our
heralded output states and the ideal squeezed Schrödinger cat states
(SOSC for odd, SESC for even), defined as
\begin{equation}
\vert\Psi_{\theta}\rangle=S(\gamma)\vert Cat\rangle_{\alpha,\theta},\label{eq:23}
\end{equation}
 where the squeezing parameter $\gamma$ to be real. The fidelity
$F=\vert\langle\Psi_{\theta}\vert\vert\Psi\rangle_{cat,m,n}\vert^{2}$
between our heralded state and this target is then maximized over
the parameters ($g$,$\alpha_{in},$$\alpha_{out},$$\gamma$), where
$\alpha_{in}$ is the input cat amplitude, $\alpha_{out}$ is the
target amplitude, and $g$ and $\gamma$ are the OPA gain and target
squeezing, respectively.

\begin{figure*}
\includegraphics[width=14cm]{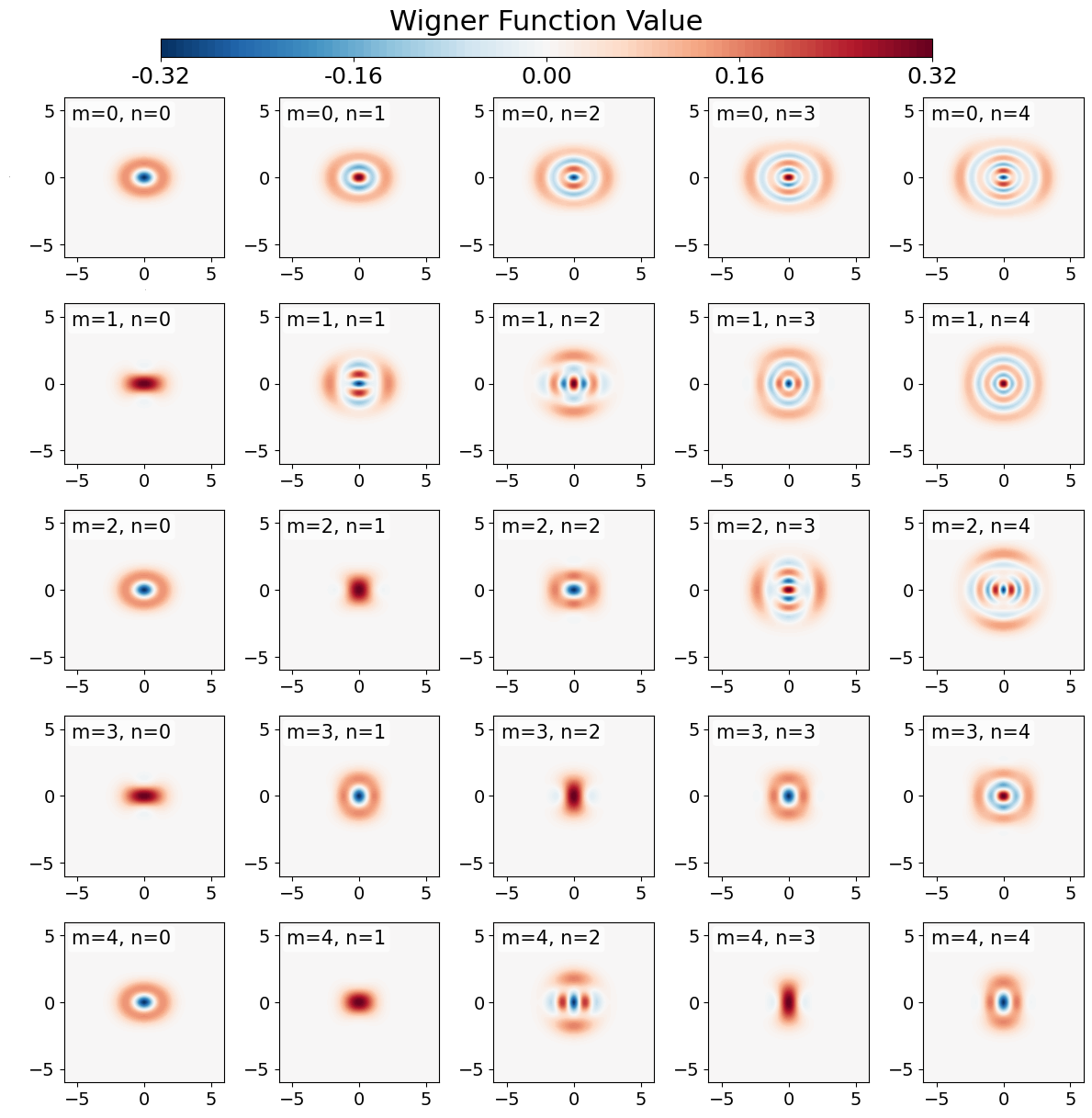}

\caption{\label{fig:5}Wigner functions of the heralded output states for odd
cat state (OCS) input for various ($m$, $n$) configurations with
$g=1.5$ and $\alpha_{in}=1.0$. The parity of these output states
is determined by $(-1)^{n+m+1}$. }
\end{figure*}

\subsection{Odd cat state input }

We first consider an odd cat state input ( $\theta=\pi$). In order
to check general feature of output signal state corresponding to the
OCS input, we plot the Wigner functions of the heralded output states
for configurations $m,n=0,1,2,3,4$ while fixing $g=1.5$ and $\alpha_{in}=1.0$
and results shown in Fig. \ref{fig:5}. It can observe that the output
state for every configuration possesses $(-1)^{n+m+1}$ parity and
have significant interference fringes characterized by pronounced
Wigner negativity. We also noticed that some configurations still
very similar to ideal squeezed cat states. To check this point, in
this subsection we mainly focus on catalytic configuration $m=n$,
where the output state retains the same parity as the input cat state.

\begin{figure}
\includegraphics[width=8.6cm,totalheight=8cm]{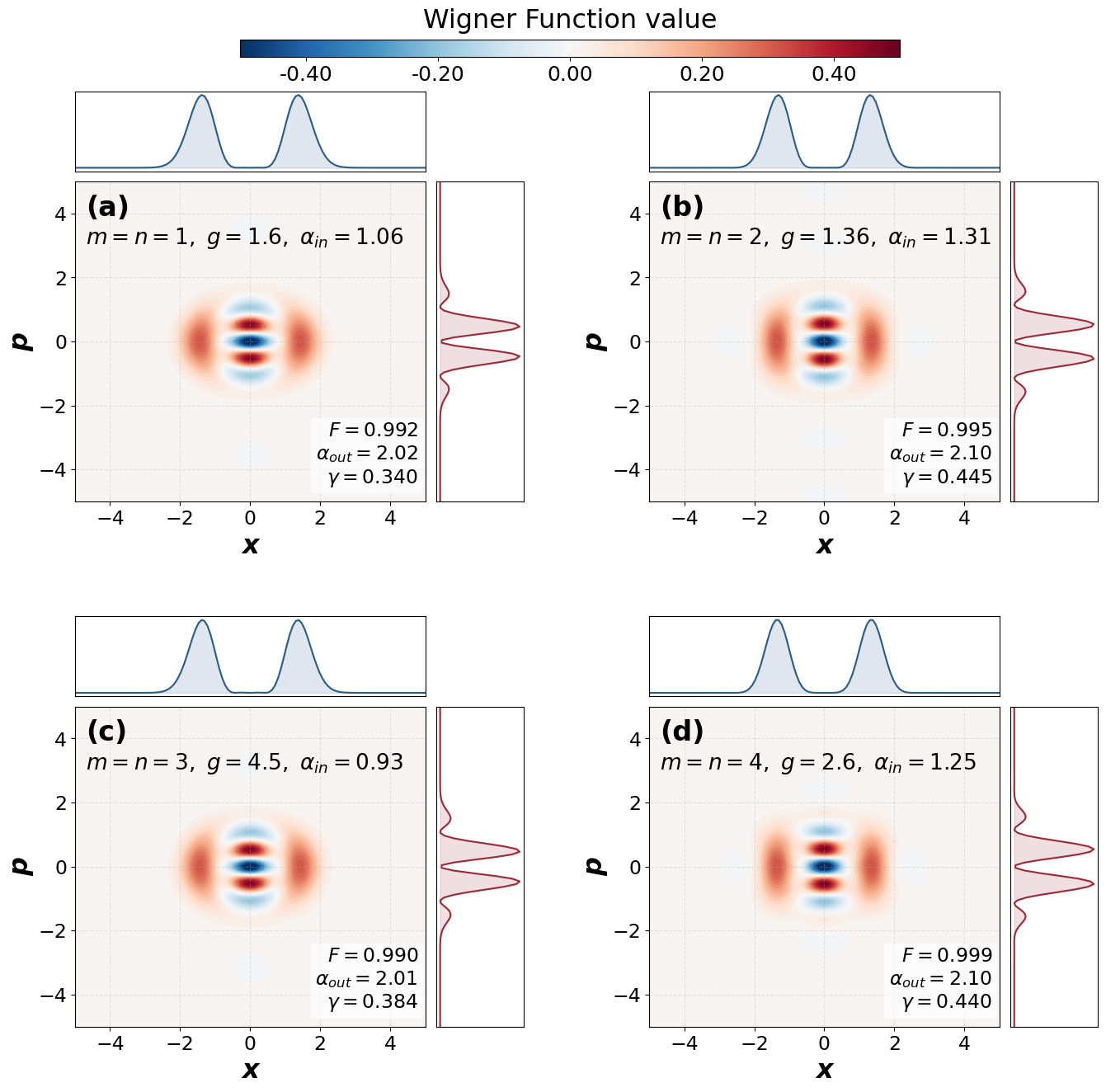}

\caption{\label{fig:6}Wigner functions of the optimized output states for
catalytic configurations $m=n=1,2,3,4$ with odd cat state (OCS) input,
showing amplified squeezed odd cat state features. In all cases the
fidelity exceeds $F>0.99$. }
\end{figure}

For the vacuum-idler case $m=n=0$, the output is simply an attenuated
OCS with reduced amplitude $\alpha/g$, as expected from the noiseless
attenuation discussed in Sec. \ref{sec:2}. By optimizing the overlap
between $\vert\Psi^{\prime}\rangle_{cat,n,n}$ ($\theta=\pi$) and
the target SOSC state $\vert\Psi_{\pi}\rangle$ we find that as the
catalysis order increases, the Wigner functions develop richer structures---negative
regions become more pronounced, and the two positive peaks move farther
apart along the $x$-axis (see Fig. \ref{fig:6}). These plots display
the characteristic features of SOSCs: a deep negative dip at the origin,
two well-separated positive peaks, and reflection symmetry about both
axes. The high similarity to the ideal SOSC targets is visually evident.
Furthermore, for input odd cat amplitudes $\alpha_{in}\lesssim1.3$,
our catalytic protocol can amplify them to large-amplitude SOSCs with
$\alpha_{out}\ge2.0$ while maintaining extremely high fidelity ($>0.99$). 

\begin{figure}
\includegraphics[width=8.6cm]{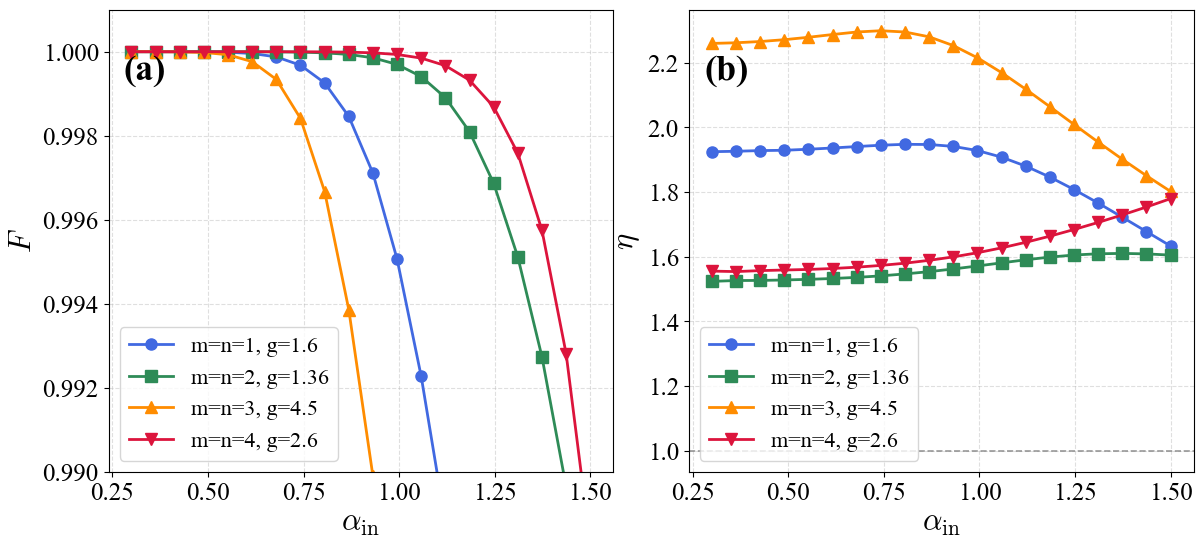}

\caption{\label{fig:7}(a) Optimal fidelity between the output state $\vert\Psi\rangle_{cat,n,n}$
and the target state $\vert\Psi_{\theta}\rangle$ as a function of
coherent amplitude of input odd cat state (OCS) for $m=n=1,2,3,4$.
(b) Corresponding amplification ratio $\eta=\alpha_{out}/\alpha_{in}$
that maximizes the fidelity in (a).}
\end{figure}

Figure \ref{fig:7} (a) shows the optimized fidelity of $\vert\Psi^{\prime}\rangle_{cat,n,n}$
($\theta=\pi$) to the target SOSC state $\vert\Psi_{\pi}\rangle$
as a function of $\alpha_{in}$ for $n=m=1,2,3,4$, while Fig. \ref{fig:7}
(b) displays the corresponding amplification ratio $\eta=\alpha_{out}/\alpha_{in}$.
From Fig. \ref{fig:7} (a), we observe that for each catalytic configuration
($n$, $n$), the fidelity exceeds $0.99$ over a finite window of
input amplitudes $\alpha_{in}$ for a specific optimized OPA gain
$g$. Specifically, for ($1$, $1$) this high-fidelity region covers
$\alpha_{in}\lesssim1.1$ ; for ($2$, $2$), it extends to $\alpha_{in}\lesssim1.43$
; for ($3$, $3$), the region is around $\alpha_{in}\lesssim0.93$;
and for ($4$, $4$), it spans $\alpha_{in}\lesssim1.48$. Within
these regions, the amplification ratio is always greater than unity,
ranging from approximately $1.52\le\eta\le2.3$. Correspondingly,
within these high-fidelity regions the output amplitude reaches values
of approximately $\alpha_{out}\lesssim2.08$ for ($1$,$1$), $\alpha_{out}\lesssim2.30$
for ($2$, $2$), $\alpha_{out}\lesssim2.09$ for ($3$, $3$), and
$\alpha_{out}\lesssim2.61$ for ($4$, $4$) {[}see Fig. \ref{fig:7}
(b) {]}. Notably, for $\alpha_{in}\lesssim1.1$, the amplification
ratio $\eta$ ranges from about $1.52$ to $2.3$, ensuring that the
output consistently falls into the large-amplitude regime $\alpha_{out}\ge2$
in a single successful trial for the specific configurations. Concrete
examples are listed in Table \ref{tab:2}. For instance, with the
simplest catalytic configuration ($1,$$1$), an input OCS with amplitude
$\alpha_{in}=1.06$, readily preparable by single-photon subtraction
\citep{2006}, is transformed into a large-amplitude SOCS with $\alpha_{out}=2.02$
at a fidelity of $0.992$ and a single-trial success probability of
$0.27\%$. This performance compares favourably with previous cat-state
amplification schemes \citep{PhysRevA.103.013710,PhysRevA.105.043713}.

Remarkably, the input amplitude $\alpha_{in}=1.06$ sits at the upper
boundary of the high-fidelity regime ( $\alpha\lesssim1.2$) for the
standard single-photon-subtracted squeezed vacuum state, which is
known to produce odd kitten states with fidelity above $0.99$ \citep{PhysRevA.70.020101,2006}.
Our ($1$, $1$) catalytic configuration takes this same type of state
and amplifies it continuously into the large-amplitude regime $\alpha_{out}\ge2.0$
while maintaining comparable fidelity---a task that conventional
photon subtraction alone cannot achieve. This demonstrates that our
protocol is not merely a state converter but a genuine amplifier of
non-Gaussian quantum resources, capable of bridging the gap between
easily producible small-amplitude cat states and the large-amplitude
states required for fault-tolerant quantum error correction \citep{R2023}. 

\begin{table}

\caption{\label{tab:2}Optimized parameters for odd cat state (OCS) amplification
under catalytic configurations $m=n=1,2,3,4$. For each configuration,
we list the OPA gain $g$, the amplitude $\alpha_{in},$the target
sate characteristics ($\alpha_{out},$squeezing $\gamma$), the success
probability $P_{trial},$and the optimized fidelity $F$.}

\begin{tabular}{|c|>{\centering}p{0.7cm}|c|c|c|c|c|}
\hline 
\multirow{2}{*}{$(m,n)$} & \multirow{2}{0.7cm}{$g$} & \multirow{2}{*}{$\alpha_{in}$} & \multicolumn{2}{c|}{Target state} & \multirow{2}{*}{$F$} & \multirow{2}{*}{$P_{trial}$}\tabularnewline
\cline{4-5}
 &  &  & $\alpha_{out}$ & $\gamma$ &  & \tabularnewline
\hline 
\hline 
\multirow{3}{*}{$(1,1)$} & \multirow{3}{0.7cm}{$1.6$} & $0.30$ & $0.58$ & $0.03$ & $1.000$ & $1.39\times10^{-3}$\tabularnewline
\cline{3-7}
 &  & $1.06$ & $2.02$ & $0.34$ & $0.992$ & $2.73\times10^{-3}$\tabularnewline
\cline{3-7}
 &  & $1.31$ & $2.31$ & $0.39$ & $0.976$ & $3.94\times10^{-3}$\tabularnewline
\hline 
\multirow{3}{*}{$(2,2)$} & \multirow{3}{0.7cm}{$1.36$} & $0.30$ & $0.46$ & $0.03$ & $1.000$ & $2.78\times10^{-4}$\tabularnewline
\cline{3-7}
 &  & $1.06$ & $1.67$ & $0.31$ & $0.999$ & $3.33\times10^{-4}$\tabularnewline
\cline{3-7}
 &  & $1.31$ & $2.11$ & $0.45$ & $0.995$ & $3.71\times10^{-4}$\tabularnewline
\hline 
\multirow{3}{*}{$(3,3)$} & \multirow{3}{0.7cm}{$4.5$} & $0.30$ & $0.68$ & $0.05$ & $1.000$ & $5.73\times10^{-7}$\tabularnewline
\cline{3-7}
 &  & $1.06$ & $2.30$ & $0.43$ & $0.980$ & $1.66\times10^{-6}$\tabularnewline
\cline{3-7}
 &  & $1.31$ & $2.56$ & $0.47$ & $0.958$ & $2.64\times10^{-6}$\tabularnewline
\hline 
\multirow{3}{*}{$(4,4)$} & \multirow{3}{0.7cm}{$2.6$} & $0.30$ & $0.47$ & $0.03$ & $1.000$ & $2.74\times10^{-7}$\tabularnewline
\cline{3-7}
 &  & $1.06$ & $1.72$ & $0.32$ & $0.999$ & $3.41\times10^{-7}$\tabularnewline
\cline{3-7}
 &  & $1.31$ & $2.24$ & $0.48$ & $0.998$ & $3.94\times10^{-7}$\tabularnewline
\hline 
\end{tabular}

\end{table}

\subsection{Even cat state input }

\begin{figure*}
\includegraphics[width=16cm]{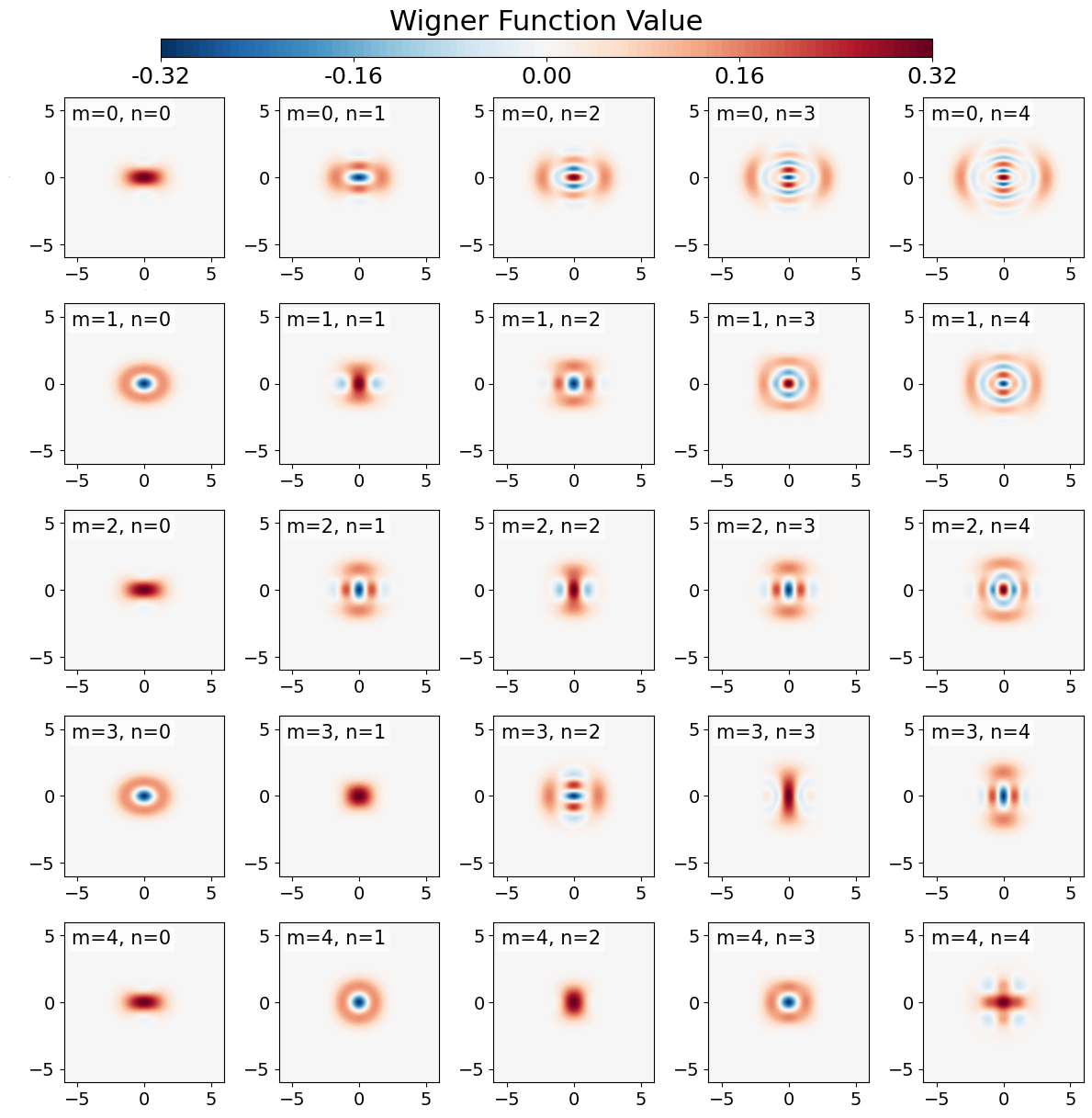}

\caption{\label{fig:8}Wigner functions of the heralded output states for even
cat state (ECS) input for various ($m$, $n$) configurations. with
$g=1.5$ and $\alpha_{in}=1.0$. The parity of these output states
is determined by $(-1)^{n+m}$.}
\end{figure*}

We now consider an even cat state input ($\theta=0$). The Wigner
functions of the heralded output states for configurations $n,m=0,1,2,3,4$
are shown in Fig. \ref{fig:8}. The parity of heralded output state
decided by $(-1)^{n+m}$ and this rule can be seen from the Wigner
function profiles present in Fig. \ref{fig:8}. For catalytic configurations
$n=m=1,2,3,...$, unlike the OCS case, ECSs have a positive central
peak at the origin and negative regions at larger phase-space radii.
As the catalysis order increases, the Wigner functions develop more
complex interference patterns, with additional negative regions emerging. 

\begin{figure}
\includegraphics[width=8.6cm]{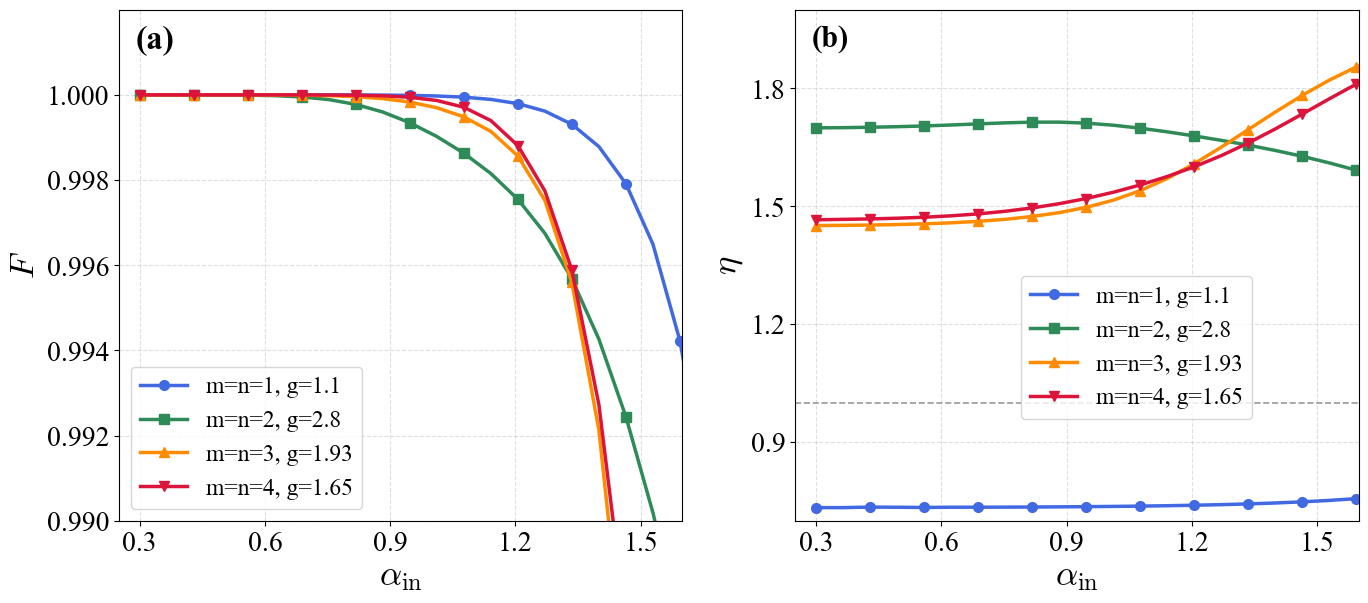}

\caption{\label{fig:10} (a) Optimized fidelity between the heralded output
state and the ideal squeezed even cat state as a function of the input
amplitude $\alpha_{in}$ for catalytic configurations $m=n=1,2,3,4$.
(b) Corresponding amplification ratio $\eta=\alpha_{out}/\alpha_{in}$
that maximizes the fidelity in (a). }
\end{figure}

Figure \ref{fig:10} shows the optimized fidelity between the heralded
output state and the ideal SESC state for catalytic configurations
$m=n=2,3,4$ as a function of the input amplitude $\alpha_{in}$.
For each configuration, the curves are obtained by optimizing over
all parameters and selecting the OPA gain $g$ that yields the highest
fidelity across the entire range of input amplitudes considered. The
fidelity remains above $0.98$ for all configurations over a broad
range of input amplitudes, with many cases achieving near-unity fidelity
( $F\ge0.999$). 

For the ECS input, the high-fidelity windows $F\ge0.99$ regions are
notably broad: ($2$, $2$) covers $\alpha_{in}\lesssim1.53$, ($3$,
$3$) extends to $\alpha_{in}\lesssim1.42$, and ($4$, $4$) spans
$\alpha_{in}\lesssim1.43$. Within these regions, the output amplitude
reaches $\alpha_{out}\lesssim2.47$, $2.50$, and $2.46$, respectively,
corresponding to a moderate amplification ratio of $1.45\le\eta\le1.8$.
Correspondingly, within these high-fidelity regions the output amplitude
reaches values of approximately $\alpha_{out}\lesssim2.47$ for ($2$,
$2$), $\alpha_{out}\lesssim2.50$ for ($3$, $3$), and $\alpha_{out}\lesssim2.46$
for ($4$, $4$). 

In Table \ref{tab:3} lists the optimized parameters and resulting
fidelities for catalytic configurations $m=n=1,2,3,4$, using the
same parameter structure as Table \ref{tab:2}. Our protocol indeed
provides an extremely good approximation to the target SESC state
when the input is ECS. For ($2$, $2$) and ($4$, $4$) with $\alpha_{in}=1.0$,
the output reaches $\alpha_{out}=1.71$ and $1.53$ with same fidelity
$F=0.999$. The corresponding success probabilities range from $10^{-5}$
down to $10^{-7}$, consistent with the trends observed for the OCS
case. 

\begin{table}
\caption{\label{tab:3}Optimized parameters ($g$, $\alpha_{in},$target state
\{$\alpha_{out},$$\gamma$\}, $P_{trial}$ and $F$) for even cat
state (ECS) amplification under catalytic configurations }

\begin{tabular}{|c|c|c|c|c|c|c|}
\hline 
\multirow{2}{*}{$(m,n)$} & \multirow{2}{*}{$g$} & \multirow{2}{*}{$\alpha_{in}$} & \multicolumn{2}{c|}{Target state} & \multirow{2}{*}{$F$} & \multirow{2}{*}{$P_{trial}$}\tabularnewline
\cline{4-5}
 &  &  & $\alpha_{out}$ & $\gamma$ &  & \tabularnewline
\hline 
\hline 
\multirow{3}{*}{$(1,1)$} & \multirow{3}{*}{$1.1$} & $0.50$ & $0.37$ & $0.015$ & $1.000$ & $4.91\times10^{-2}$\tabularnewline
\cline{3-7}
 &  & $1.00$ & $0.74$ & $0.061$ & $1.000$ & $3.64\times10^{-2}$\tabularnewline
\cline{3-7}
 &  & $1.50$ & $1.13$ & $0.149$ & $0.997$ & $1.67\times10^{-2}$\tabularnewline
\hline 
\multirow{3}{*}{$(2,2)$} & \multirow{3}{*}{$2.8$} & $0.50$ & $0.85$ & $0.070$ & $1.000$ & $1.62\times10^{-5}$\tabularnewline
\cline{3-7}
 &  & $1.00$ & $1.71$ & $0.246$ & $0.999$ & $4.43\times10^{-5}$\tabularnewline
\cline{3-7}
 &  & $1.50$ & $2.43$ & $0.397$ & $0.991$ & $9.49\times10^{-5}$\tabularnewline
\hline 
\multirow{3}{*}{$(3,3)$} & \multirow{3}{*}{$1.93$} & $0.50$ & $0.73$ & $0.016$ & $1.000$ & $3.33\times10^{-6}$\tabularnewline
\cline{3-7}
 &  & $1.00$ & $1.51$ & $0.262$ & $0.999$ & $5.70\times10^{-6}$\tabularnewline
\cline{3-7}
 &  & $1.50$ & $2.70$ & $0.661$ & $0.981$ & $8.60\times10^{-6}$\tabularnewline
\hline 
\multirow{3}{*}{$(4,4)$} & \multirow{3}{*}{$1.65$} & $0.50$ & $0.74$ & $0.076$ & $1.000$ & $3.85\times10^{-7}$\tabularnewline
\cline{3-7}
 &  & $1.00$ & $1.53$ & $0.309$ & $0.999$ & $6.41\times10^{-7}$\tabularnewline
\cline{3-7}
 &  & $1.50$ & $2.63$ & $0.688$ & $0.984$ & $8.84\times10^{-7}$\tabularnewline
\hline 
\end{tabular}
\end{table}

\begin{figure}
\includegraphics[width=8.6cm]{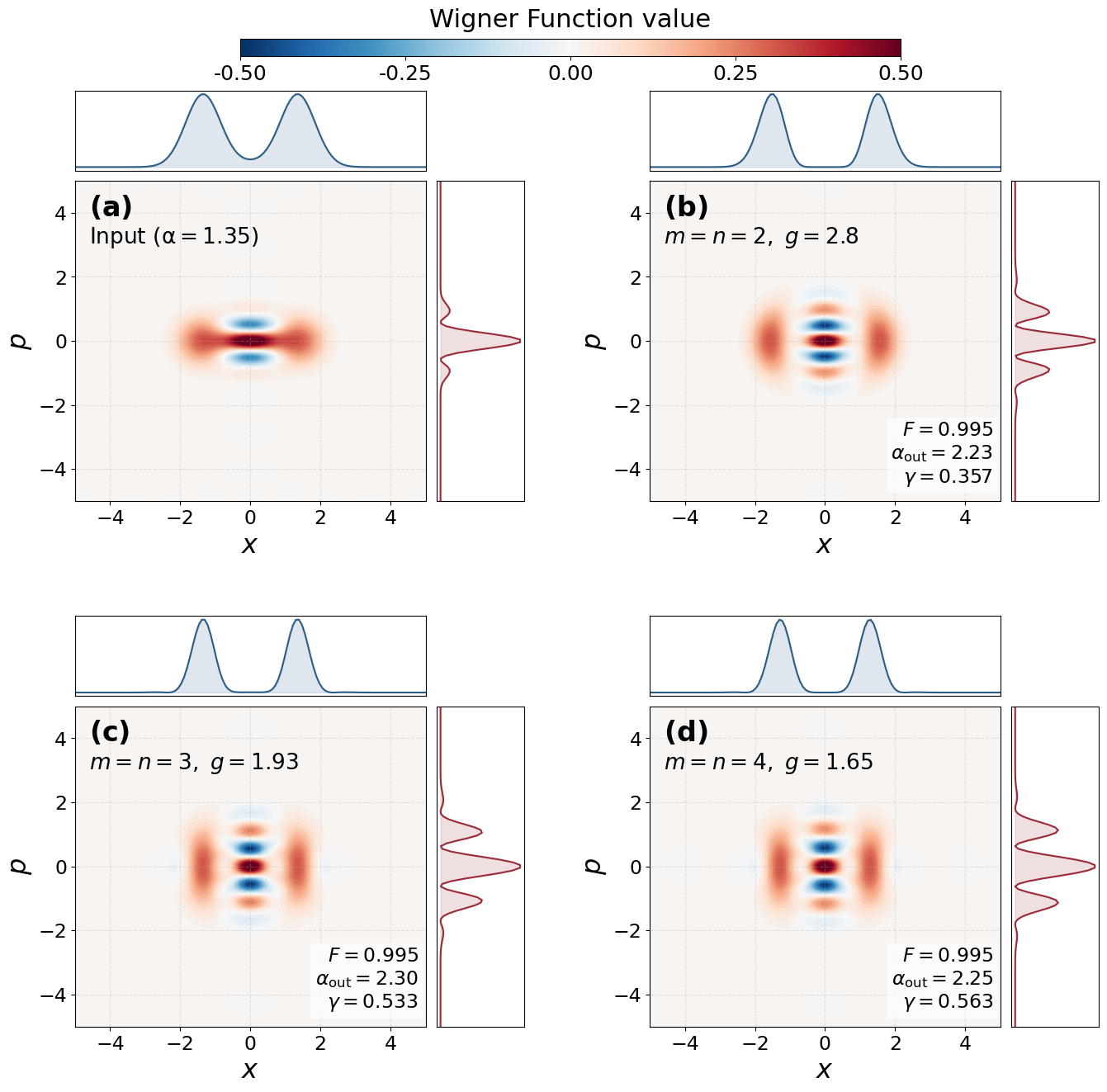}

\caption{\label{fig:9}Wigner functions of the optimized output states for
catalytic configurations $m=n=2,3,4$ with even cat state (ECS) input,
showing amplified squeezed even cat state features with $F=0.995.$ }
\end{figure}

To further confirm our claim, we present in Fig. \ref{fig:9} the
Wigner functions of the output states for catalytic configurations
($2$, $2$), ($3$, $3$) and ($4$, $4$), corresponding to an ECS
input with amplitude $\alpha_{in}=1.35$. All three configurations
produce excellent approximations to the target SESC states, with amplified
coherent amplitudes $\alpha_{out}=2.23$, $2.30$, and $2.25$, respectively,
while maintaining exceptionally high fidelity $F=0.995$. The corresponding
success probabilities are on the order of $10^{-5}$ for ($2$, $2$)
, $10^{-6}$ for ($3$, $3$), and $10^{-7}$ for ($4$, $4$). These
Wigner plots provide direct visual evidence that our protocol reliably
generates large-amplitude squeezed cat states with high fidelity across
different catalysis orders.

We note an interesting contrast with the OCS case. For ECS inputs,
the ($1,1$) configuration does not amplify the state but instead
acts as a noiseless attenuator, yet its performance remains remarkably
good---for example, reducing $\alpha_{in}=1.0$ to $\alpha_{out}=0.74$
with fidelity $F=1.000$ {[}see Table \ref{tab:3}{]}. This attenuation
behavior is also evident in Fig. \ref{fig:10} (b), where the amplification
ratio $\eta=\alpha_{out}/\alpha_{in}$ remains below unity across
the entire range of $\alpha_{in}$ considered. This confirms that
the ($1,$$1$) catalysis acts as a genuine attenuator for even cat
inputs, in stark contrast to its amplifying behavior for OCS inputs.

A plausible explanation for this parity-dependent behavior lies in
the explicit form of the ($1,$$1$) heralded state 

\begin{equation}
\vert\Psi\rangle_{\theta,1,1}\propto\frac{1}{g^{2}}\vert Cat\rangle_{\alpha/g,\theta}-\left(1-\frac{1}{g^{2}}\right)a^{\dagger}a\vert Cat\rangle_{\alpha/g,\theta},
\end{equation}
where $\theta=0$ ($\theta=\pi$) for ECS (OCS) input case. For an
ECS input, the operator $a^{\dagger}a$ removes the vacuum component
of the state, reducing the relative weight of low-lying Fock components
which destroy the cat state structure. If we consider the medium OPA
gain cases, the first term is dominated and for both input cases can
give noiseless attenuation for corresponding cat state inputs. To
confirm that for ECS input this is indeed attenuation rather than
amplification, we have scanned the full range of allowed OPA gain
$g$ for the ($1$, $1$) configuration with ECS inputs. The fidelity
to the target SESC state remains very low for all values of $g$ except
in the specific parameter region corresponding to the noiseless attenuation.
However, for OCS input the result is different. Since the OCS has
no vacuum component, the operator $a^{\dagger}a$ effectively increases
the mean photon number of the heralded sate $\vert\Psi\rangle_{\pi,1,1}$
rather than destroy the inherent structure of the OCS input. By adjusting
the OPA gain $g$ to larger values, the second term is dominated and
could give amplified SOSC state {[}see Fig. \ref{fig:7} and Table
\ref{tab:2}{]}. This provides a clear physical picture of why the
($1,$$1$) catalysis attenuates ECS while amplifying OCS.

\subsection{Non-catalytic configurations}

In addition to the catalytic cases, we have also investigated non-catalytic
configurations ($m\neq n$) for cat state amplification. Here, the
parity of our approximated cat sates are not preserved, and determined
by $(-1)^{m+n}$ and $(-1)^{m+n+1}$ for ECS and OCS inputs, respectively. 

\begin{figure}
\includegraphics[width=8.6cm]{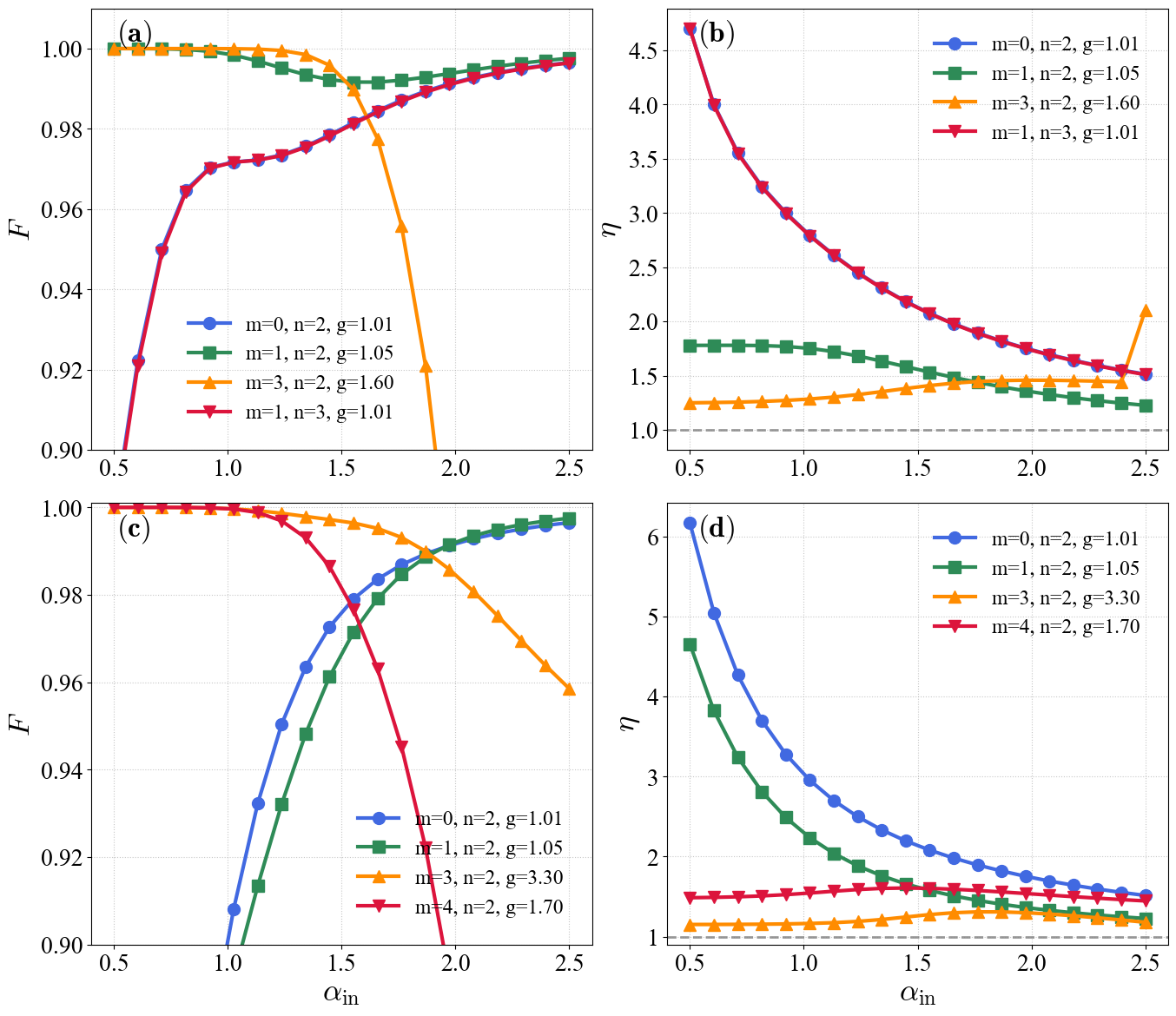}

\caption{\label{fig:11}Optimized fidelity (left column) and amplification
ratio (right column) for non-catalytic configurations ($m\protect\neq n$):
(a,b) for for even cat state (ECS) inputs with configurations ($0$,
$2$), ($1,$$2$), ($1,$$3$) and ($3,\text{2}$); (c,d) for odd
cat inputs with configurations ($0$, $2$), ($1$, $2$), ($3$,
$2$) and ($4$, $2$). }

\end{figure}

Figure \ref{fig:11} shows the optimized fidelity as a function of
the input amplitude $\alpha_{in}$ for several non-catalytic configurations,
including ($0$, $2$), ($1$, $2$), ($1$, $3$) and ($3$, $2$)
for ECS inputs (Fig.\ref{fig:11} (a) and (b)), and ($0$, $2$),
($1$, $2$), ($3$, $2$) and ($4$, $2$) for OCS inputs (Fig.\ref{fig:11}
(c) and (d)). For each configuration, the OPA gain $g$ is optimized
to maintain high fidelity across the considered range of input amplitudes.
For both inputs, the amplification ratio $\eta=\alpha_{out}/\alpha_{in}$
is consistently greater than unity, ensuring that the output falls
into the large-amplitude regime for appropriate input parameters {[}see
Fig.\ref{fig:11} (b) and (d){]}. 

Table \ref{tab:4} presents the optimized parameters and fidelities
for representative non-catalytic configurations. For ECS inputs, the
($1,$$3$) configuration achieves a fidelity of $0.991$ with $\alpha_{out}=3.45$
and $\gamma=0.26$, at a success probability of $1.13\times10^{-3}$.
The reference ($0$ , $2$) configuration---previously studied in
the context of photon addition \citep{Park:16}---achieves a comparable
fidelity of $0.991$ with a similar output amplitude and success probability.
The catalytic configurations ($2$, $2$) , ($3$, $3$) and ($4$,
$4$) from Table \ref{tab:3} achieves comparable fidelity but with
a lower success probability. This illustrates the trade-off between
catalytic advantages (parity preservation, idler restoration) and
generation rate.

For OCS inputs, the non-catalytic configurations exhibit distinct
trade-offs. The ($1$, $2$) configuration achieves a fidelity of
$0.991$ with a practical success probability of $2.2\times10^{-2}$,
amplifying $\alpha_{in}=1.97$ to $\alpha_{out}=2.69$. While the
($3$, $2$) configuration marginally improve the fidelity to $0.993$,
amplifying $\alpha_{in}=1.76$ to $\alpha_{out}=2.31$. This comes
at a prohibitive cost---the success probability plummets to $5.52\times10^{-6}$
(a four-orders-of-magnitude reduction), rendering it experimentally
inefficient for most purposes. Interestingly, the reference ($0$,
$2$) configuration yields the largest output amplitude among the
three which amplifying $\alpha_{in}=1.97$ to $\alpha_{out}=3.46$
with a moderate success probability of $4.59\times10^{-3}$ and a
fidelity of $0.991$. This highlights that the ($0$, $2$) scheme
serves as an \textquotedbl amplitude-maximizing\textquotedbl{} amplifier,
whereas the ($1$, $2$) case provides the best overall balance between
fidelity, rate, and operable bandwidth. Collectively, these results
demonstrate that non-catalytic configurations offer flexible trade-offs---achieving
comparable or even superior performance for specific targets at the
expense of losing the parity preservation and idler restoration unique
to the catalytic case.

Remarkably, the ($1$, $2$) non-catalytic configuration with ECS
inputs stands out as one of the best-performing cases in our entire
protocol. Over a wide input range $\alpha_{in}\in\left[0.5,2.5\right]$,
the fidelity to the target SOSC state consistently exceeds $0.992$,
while the amplification ratio ranges from $\eta\approx1.78$ at small
amplitudes to $\eta\approx1.23$ at larger ones, and the success probability
remains between $1.2\%$ and $2.2\%$ {[}see Fig. \ref{fig:11} (a)
and (b){]}. For example, with the ($1$, $2$ ) non-catalytic configuration,
an ECS input of amplitude $\alpha_{in}=1.18$ is transformed into
a target SOSC state with amplitude $\alpha_{out}=2.02$ and squeezing
$\gamma=0.23$, while maintaining a high fidelity of $0.996$ and
a success probability of $1.74$\% {[}see Table \ref{tab:4}{]}. This
means that small even kittens can be amplified to large-amplitude
SOCSs with near-perfect fidelity across a broad parameter regime.
To the best of our knowledge, such high-fidelity amplification of
ECSs to OCSs---accompanied by a parity flip---has not been reported
in previous cat-state amplification schemes. This parity-flipping
amplification capability, combined with the parity-preserving catalysis
demonstrated earlier, establishes our OPA platform as a versatile
tool for controlled cat-state engineering with tunable parity outcomes.

\begin{table*}
\caption{\label{tab:4}Optimized parameters ($g$, $\alpha_{in},$target state
\{$\alpha_{out},$$\gamma$, parity\}, $P_{trial}$ and $F$) for
cat state amplification under non-catalytic configurations ($m\protect\neq n$),
for both even cat state (ECS) and odd cat state (OCS) inputs.}

\begin{tabular}{|c|c|c|>{\centering}p{1.1cm}|>{\centering}p{1.1cm}|>{\centering}p{1.1cm}|c|>{\centering}p{1.1cm}|c|}
\hline 
\multirow{2}{*}{Input } & \multirow{2}{*}{$(m,n)$} & \multirow{2}{*}{$g$} & \multirow{2}{1.1cm}{$\ \ \alpha_{in}$} & \multicolumn{3}{c|}{Target state} & \multirow{2}{1.1cm}{$\ \!\ F$} & \multirow{2}{*}{$\ \ P_{trial}$}\tabularnewline
\cline{5-7}
 &  &  &  & $\alpha_{out}$ & $\gamma$ & Parity &  & \tabularnewline
\hline 
\hline 
\multirow{4}{*}{ECS} & $(1,2)$ & $1.05$ & $1.18$ & $2.02$ & $0.23$ & Odd & $0.996$ & $1.74\times10^{-2}$\tabularnewline
\cline{2-9}
 & $(1,3)$ & $1.01$ & $1.97$ & $3.45$ & $0.26$ & Even & $0.991$ & $1.13\times10^{-3}$ \tabularnewline
\cline{2-9}
 & $(0,2)$ & $1.01$ & $1.97$ & $3.46$ & $0.26$ & Even & $0.991$ & $4.59\times10^{-3}$\tabularnewline
\cline{2-9}
 & $(3,2)$ & $1.60$ & $1.45$ & $2.0$0 & $0.49$ & Odd & $0.996$ & $1.34\times10^{-5}$\tabularnewline
\hline 
\multirow{4}{*}{OCS} & $(1,2)$ & $1.05$ & $1.97$ & $2.69$ & $0.20$ & Even & $0.991$ & $2.20\times10^{-2}$\tabularnewline
\cline{2-9}
 & $(0,2)$ & $1.01$ & $1.97$ & $3.46$ & $0.25$ & Odd & $0.991$ & $4.59\times10^{-3}$\tabularnewline
\cline{3-9}
 & $(3,2)$ & $3.30$ & $1.76$ & $2.31$ & $0.35$ & Even & $0.993$ & $5.52\times10^{-6}$\tabularnewline
\cline{2-9}
 & $(4,2)$ & $1.70$ & $1.34$ & $2.15$ & $0.51$ & Odd & $0.993$ & $2.59\times10^{-7}$\tabularnewline
\hline 
\end{tabular}
\end{table*}

\subsection{Discussion of amplification performance}

Our protocol amplifies input cat states to large-amplitude squeezed
cat states with high fidelity ($F\ge0.99$) and success probabilities
ranging from $10^{-2}$ to $10^{-7}$ across various catalytic and
non-catalytic configurations. ECS inputs yield higher fidelities than
OCS inputs for the same catalysis order, particularly at small amplitudes,
but exhibit less pronounced amplification ( $\alpha_{out}/\alpha_{in}$).
This trade-off reflects the larger vacuum component of ECSs, which
improves target-state overlap but limits amplitude gain. 

To generate large-amplitude cat states, numerous theoretical and experimental
schemes have been proposed, including photon subtraction \citep{Park:16,PhysRevA.111.043704,2025K,Wang:25,PhysRevA.105.043713,RN87,PhysRevA.103.013710}
and photon catalysis\citep{Eaton_2019} and postselected measurement
\citep{Yao_2026,Npj2026} schemes were conducted. However, many of
these approaches face significant experimental overhead. Multiphoton
(thee-photon) subtraction schemes produce amplitudes $\alpha\approx1.6$-$1.7$
with success probability $\sim10^{-9}$ \citep{PhysRevA.82.031802,Yukawa:13},
while Song et al.\citep{PhysRevA.105.043713} predict $\alpha_{out}\approx2.51$
with success probability $\sim10^{-7}$. The homodyne breeding protocol
of Sychev et al. \citep{RN87} achieves $\alpha_{out}\approx1.85$
with success probability $10^{-8}$ via two-photon detection with
additional homodyne conditioning. More recently, Erkilic et al. \citep{2025K}
demonstrated an OPA-based scheme that generates large-amplitude squeezed
cat states ($\alpha\approx4.9-5.7$, $F\approx0.99$ ) using only
Gaussian inputs (squeezed vacuum) and low per-round photon detection
($2$-$3$photons), but requires $5-6$ iterative rounds with an optical
switch, leading to cumulative photon detections of $12-15$ photons
and overall success probabilities of $10^{-4}$- $10^{-6}$, limited
by switching loss. While their approach eliminates the need for non-Gaussian
seed states, the iterative nature introduces experimental complexity
and loss accumulation.

For high-success-rate generation, schemes based on two squeezed-state
inputs interfering on a beam splitter with heralded photon detection
have been proposed. Luo and Mahmoodian \citep{t96h-488y} achieve
$F\ge0.99$, $\alpha_{out}>2$, and success probabilities $\sim0.5$
with single-photon detection, but require strong squeezing and a dual-input
interferometric setup with stringent phase synchronization and additional
non-Gaussian resource preparation. The generalized photon subtraction
(GPS) scheme \citep{PhysRevA.103.013710} generates $\alpha_{out}=\sqrt{10}\approx3.16$
with $F\approx0.997$ via $10$-photon detection, yet demands extreme
squeezing and high-order photon-number resolution.

Our OPA-based protocol achieves output amplitudes comparable to those
of previous schemes, but with a different experimental cost structure.
As shown in Table \ref{tab:2}-\ref{tab:4}, the generated states
exhibit $\alpha_{out}$$\ge2.0$, proper squeezing, and $F\ge0.99$.
Notably, the ($0$, $2$) and ($1$, $3$) configurations------requiring
only $2$- or $3$- photon detection in a single pass---yield $\alpha_{out}\approx3.46$,
$F=0.991$, and success probability on the order of $10^{-3}$. These
amplitudes and fidelities are comparable to the GPS and Erkilic's
schemes, while operating at 100 MHz also enables Million counts per
second(Mcps)-level generation rates.

The cost of the present scheme is that for some configurations it
requires a seed cat state with larger amplitude, such as $\alpha_{in}\approx1.97$.
However, such seed states are readily producible with current laboratory
technology using various established methods \citep{PhysRevLett.115.023602,Yao_2026,Npj2026,turek2026},
and the gain operations used in this work are experimentally undemanding.
Critically, $1$-$3$ photon detection is experimentally mature, with
PNRD efficiency approaching unity at low photon numbers. In contrast,
iterative schemes and high-photon-number (e.g., $10$-photon) resolution
suffer from cumulative loss, synchronization overhead, and significant
efficiency degradation.

Importantly, the amplified cat states produced by our protocol can
themselves serve as seeds for subsequent amplification rounds. For
example, a small-amplitude cat state ($\alpha_{in}\approx1$)---readily
generated by standard single-photon subtraction---can be amplified
via the ($1$, $1$) catalytic configuration to $\alpha_{out}\approx2.0$.
We note that a SOCS with $\alpha_{out}\approx2$ and $F=0.992$ can
also be generated from a squeezed vacuum input using the ($1$, $2$)
configuration \citep{turek2026}. These outputs can then be fed into
a second amplification round using, for instance, the ($1$, $2$)
non-catalytic configuration, yielding $\alpha_{out}\approx3.46$.
Assuming a one-time seed preparation success probability of $\sim10^{-2}$
and a single-round OPA heralding probability of $\sim10^{-3}$, the
total success probability for two rounds is on the order of $10^{-5}$
to $10^{-6}$. This is comparable to that of iterative OPA schemes
\citep{2025K} and higher than multi-photon subtraction schemes \citep{PhysRevA.103.013710,PhysRevA.105.043713},
yet our protocol avoids the need for extreme squeezing, high-order
photon detection, or optical switching.

This approach thus offers a flexible and complementary route to large-amplitude
squeezed cat states, using moderate resources instead of extreme squeezing,
high-order detection, or iterative breeding. By selecting the appropriate
catalytic or non-catalytic configuration, the protocol can be tailored
to suit specific application needs while maintaining high generation
rates suitable for fault-tolerant quantum computing \citep{PhysRevA.106.022431,jxbx-75m9,97yt-nzg2,hr5f-lvy7}.

\section{\label{sec:5}Photon loss}

Since the squeezing operation can reduce the average photon number
of cat states, the squeezed cat states survive longer than usual cat
states in a lossy environment. To evaluate the practical feasibility
of our protocol, we model photon loss using the Lindblad master equation 

\begin{equation}
\frac{d\rho(t)}{dt}=\kappa\left(2a\rho(t)a^{\dagger}-a^{\dagger}a\rho(t)-\rho(t)a^{\dagger}a\right),\label{eq:38}
\end{equation}
 where $\kappa$ the photon loss and $a$ ($a^{\dagger}$) is the
annihilation (creation) operator of the signal mode. This master equation
is equivalent to coupling the signal mode to a vacuum environment
through a fictitious beam splitter with transmissivity $e^{-\kappa t}$,
where $t$ is the interaction time. The solution to Eq. (\ref{eq:38})
maps the initial state $\rho_{0}$ to the lossy state $\rho_{\kappa}=\mathcal{L_{\kappa}}(\rho_{0})$,
with the average photon number scaling as $\langle n\rangle\rightarrow e^{-\kappa t}\langle n\rangle$. 

In this section we also illustrate the effects of photon loss on the
negativity and complexity of our heralded states. The Winger negativity
of a quantum state $\rho$ is defined as \citep{Kenfack2004}
\begin{equation}
\mathcal{N}=\int_{\mathbb{C}}\vert W(\alpha\vert\rho)\vert\frac{d^{2}\alpha}{\pi}-1,\label{eq:18}
\end{equation}
 where $W(\alpha\vert\rho)$ is the Wigner function of the state $\rho$.
This quantity characterizes the volume of the negative part of the
Wigner function in phase space. It serves as a strong indicator of
non\nobreakdash-classicality: a non\nobreakdash-zero value of $\mathcal{N}$
signals genuine quantum interference. However, Wigner negativity does
not very accurately capture structural features such as squeezing
or high photon number when the Wigner function remains non\nobreakdash-negative
(e.g., for squeezed vacuum state).

In contrast, the complexity $\mathcal{C}(\rho)$ introduced by Tang
et al. \citep{Tang_2025} is defined via the always\nobreakdash-positive
Husimi function $Q(\alpha\vert\rho)=\langle\alpha\vert\rho\vert\alpha\rangle$:
\begin{equation}
\mathcal{C}(\rho)=e^{S_{W}(\rho)-1}I(\rho)\label{eq:19-1}
\end{equation}
 with the Wehrl entropy 
\begin{equation}
S_{W}(\rho)=-\int Q(\alpha\vert\rho)\ln Q(\alpha\vert\rho)\frac{d\alpha^{2}}{\pi}
\end{equation}
and the Fisher information 
\begin{equation}
I(\rho)=\frac{1}{4}\int\frac{\vert\vert\nabla Q(\alpha\vert\rho)\vert\vert^{2}}{Q(\alpha\vert\rho)}\frac{d\alpha^{2}}{\pi}.
\end{equation}
The quantity $\mathcal{C}(\rho)$ captures the configurational trade\nobreakdash-off
between the spread (disorder) and localization (order) of the state
in phase space. It is minimal ( $\mathcal{C}=1$) for all displaced
thermal states (including vacuum and coherent states) and increases
with squeezing and photon number. Importantly, complexity can be large
even when the Wigner function is completely non\nobreakdash-negative
(e.g., SV), thus revealing a form of structural richness that is independent
of negativity. Wigner negativity and complexity can provide complementary
insights to describe the phase space description of a given state
in detail. 

Since, the photon loss mechanism is same for all configuration, in
below we only focus one some typical signal output states characterized
by the configurations ($m$, $n$ ) which listed in Table \ref{tab:1}
-- Table \ref{tab:4}.

\begin{figure}
\includegraphics[width=8.6cm]{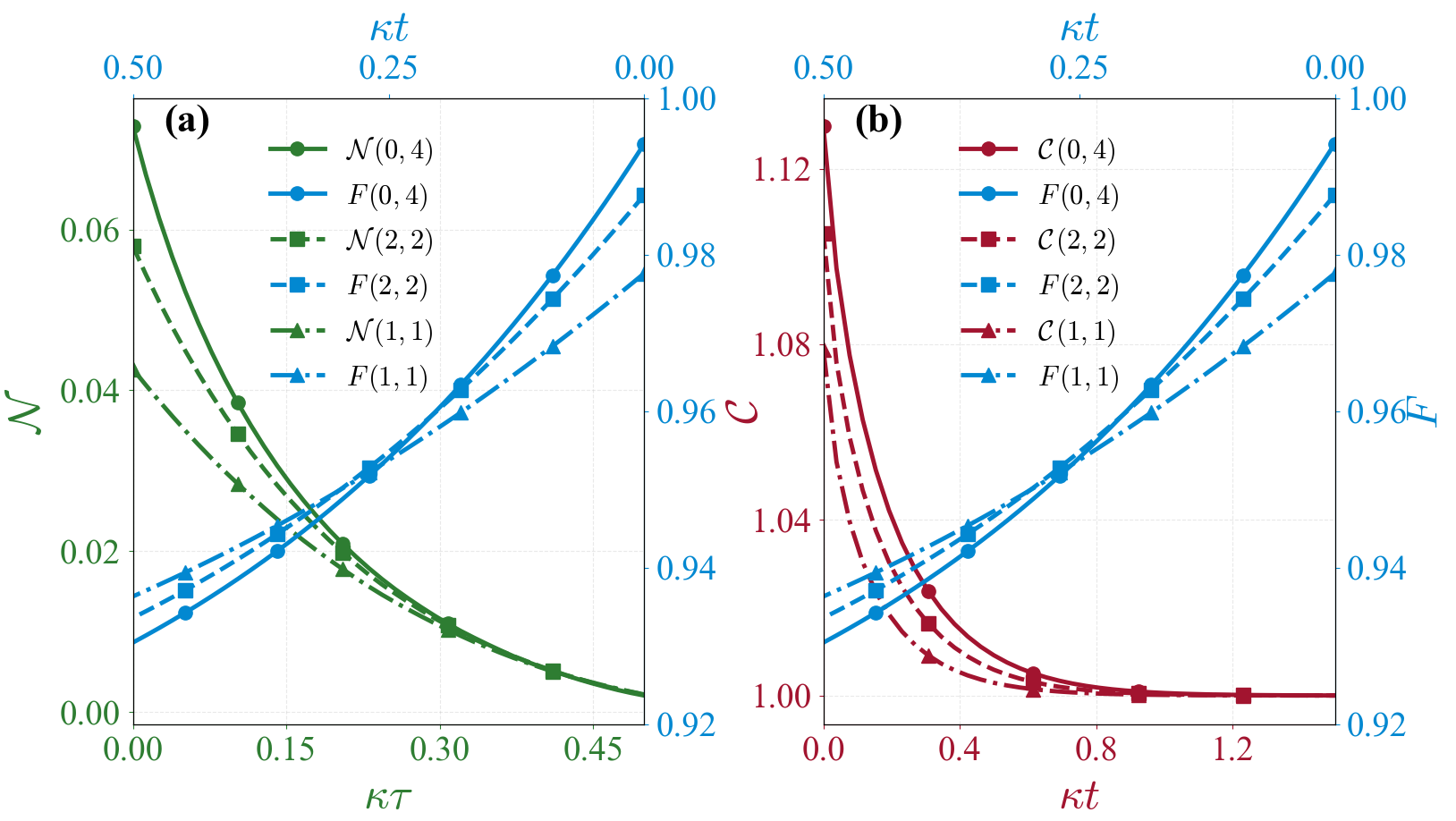}

\caption{\label{fig:14}Changes of the fidelity $F$, Wigner negativity $\mathcal{N}$,
and complexity $\mathcal{C}$ of the heralded cubic-phase-state outputs
for configurations ($0$, $4$), ($2$, $2$), and ($1$, $1$) as
functions of the photon loss rate $\kappa t$.}
\end{figure}

We choose the ($0$, $4$), ($1,1$) and ($2$, $2$) configurations
to analyze the robustness of our approximated CPSs against pure photon
loss arising from experimental imperfections; the numerical results
are shown in Fig. \ref{fig:14}. Our heralded output states are all
good approximations to the ideal CPSs, with fidelities $F=0.997$
for ($0$, $4$); $F=0.975$ for $(1,\text{1})$ and $F=0.993$ for
($2$, $2$) {[}see Table \ref{tab:1}{]}.

In Fig. \ref{fig:14}, we plot optical fidelity $F$, negativity $\mathcal{N}$
and complexity $\mathcal{C}$ of our signal output states $\vert\Psi\rangle_{coh,m,n}$
(see Eq. \ref{eq:co2}) as a function of the single-photon loss rate
$\kappa t$. The robustness against photon loss follows the ordering
$(0,4)>(2,\text{2})>(\text{1},\text{1})$. Remarkably, the fidelity
to the ideal CPS remains above $0.90$ even for strong loss rates
up to $\kappa t=0.50$, indicating that our generated states exhibit
strong resilience to single-photon loss---a favorable feature for
their use in noisy quantum information processing environments.

From the negativity curves Fig. \ref{fig:14} (a), we observe that
the negative regions of the Wigner functions for the selected configurations
are small but nonzero. This implies that these states do not exhibit
pronounced quantum interference fringes, consistent with their relatively
modest non-classicality as measured by negativity. However, as shown
in Fig. \ref{fig:14} (b), the initial complexities ($\kappa t=0$)
are significant. As photon loss increases, the negativity gradually
diminishes and vanishes for $\kappa t>0.50$, indicating that the
states become Gaussian-like in terms of their Wigner-function profile.
Nevertheless, the complexity $\mathcal{C}$ persists up to $\kappa t=1$.
This demonstrates that the structural complexity of our generated
states can survive strong photon loss rates, even after the negative
regions in phase space have entirely disappeared.

\begin{figure}
\includegraphics[width=8.6cm]{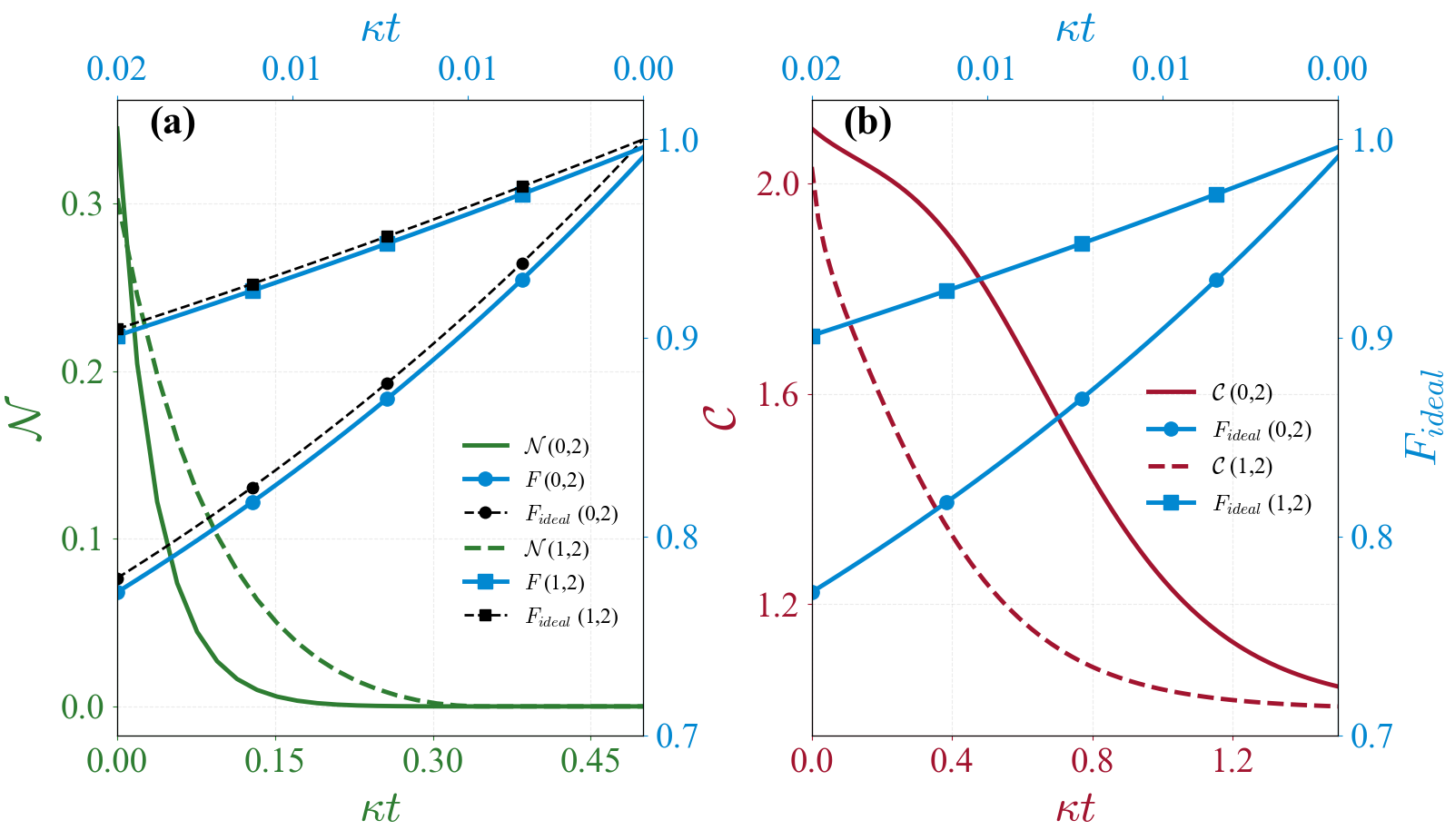}

\caption{\label{fig:12}Changes of the fidelity $F$, Wigner negativity $\mathcal{N}$,
and complexity $\mathcal{C}$ of the heralded amplified cat states
for even cat input under configurations ($0$, $2$) and ($1$, $2$)
as functions of the photon loss rate$\kappa t$. The black curves
represent the fidelity $F_{ideal}=\vert\langle\Psi_{\theta}\vert\Phi_{\theta}\rangle\vert^{2}$
between the ideal squeezed cat state and its single-photon-loss counterpart
for the same parameters (see main text for details)}

\end{figure}

\begin{figure}
\includegraphics[width=8.6cm]{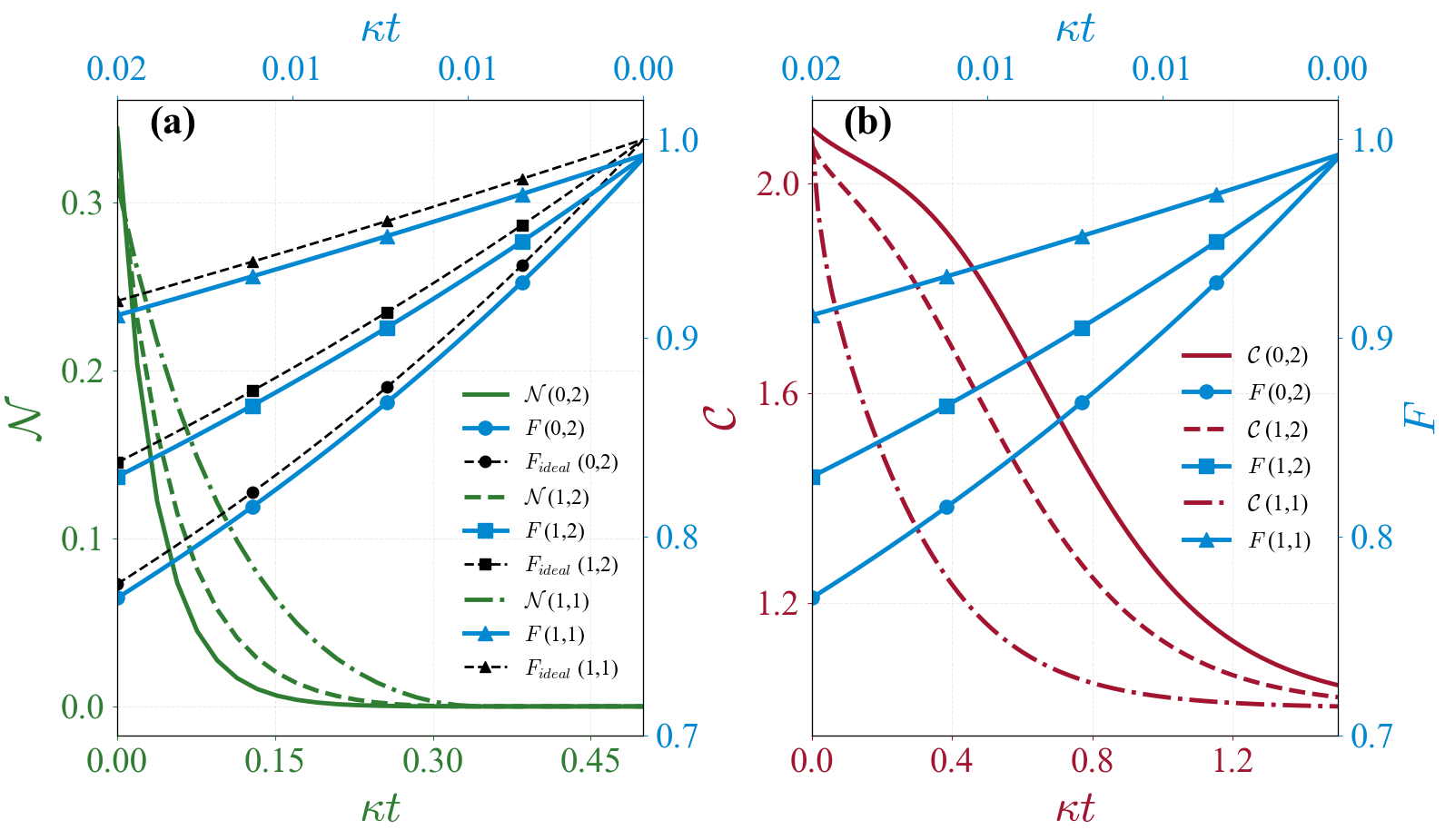}

\caption{\label{fig:13}Changes of the fidelity $F$, Wigner negativity $\mathcal{N}$,
and complexity $\mathcal{C}$ of the heralded amplified cat states
for odd cat input under configurations ($0$, $2$), ($1$, $1$)
and ($1$, $2$) as functions of the photon loss rate$\kappa t$.
The black curves represent the fidelity $F_{ideal}$ between the ideal
squeezed cat state and its single-photon-loss counterpart for the
same parameters.}
\end{figure}

In Fig. \ref{fig:12} and \ref{fig:13}, we plot the variations of
the system quantities as functions of the single-photon loss rate
$\kappa t$ for the ($1$, $2$) and ($0$, $2$) configurations with
an ECS input, and for ($1,$$1$), ($1$, $2$) and ($0$, $2$) configurations
with an OCS input, respectively. The heralded output states corresponding
to these configurations provide excellent approximations to the ideal
squeezed cat states of the corresponding parity, with output amplitudes
larger than $2.0$ and extremely high fidelities ($F\ge0.99$) {[}see
Table \ref{tab:2} and Table \ref{tab:4}{]}. 

The robustness of these good approximations against photon loss, characterized
by the fidelity $F$ and the Wigner negativity $\mathcal{N}$, follows
the ordering $(1,\text{2})>(\text{0},\text{2})$ for ECS input, and
$(1,1)>(1,2)>(0,2)$ for OCS input. Among these, the ($1$, $2$)
and ($1,1$) configurations---which effectively amplify small kitten
states to large-amplitude ($\alpha_{out}\ge2.0$) squeezed cat states---exhibit
excellent robustness, maintaining the fidelity above $0.9$ for $\kappa t\le0.02$.
Although this loss window is narrower than that observed for the CPS
cases {[}see Fig. \ref{fig:14}{]}, it is nevertheless a noteworthy
performance for cat states, given their more complex interference
structures and higher sensitivity to decoherence. 

The ordering of the complexity $\mathcal{C}$ as a function of $\kappa t$
is $(0,\text{2})>(\text{1},\text{2})$ for ECS inputs, and $(0,2)>(1,2)>(1,1)$
for OCS inputs. As shown in Fig. \ref{fig:12} and \ref{fig:13},
these good approximations to amplified cat states exhibit relatively
strong complexity $\mathcal{C}$ and significantly more pronounced
negativity $\mathcal{N}$ compared to the CPS cases presented in Fig.
\ref{fig:14}. Interestingly, the complexity of these states persists
even under very strong single-photon loss rates, long after the corresponding
Wigner negativity has completely vanished. 

For reference, we also plot the fidelity $F_{ideal}=\vert\langle\Psi_{\theta}\vert\Phi_{\theta}\rangle\vert^{2}$
between the ideal squeezed Schrödinger cat state $\vert\Psi_{\theta}\rangle$
{[}see Eq. (\ref{eq:23}){]} and the squeezed Schrödinger cat state
after a single-photon loss channel$,\vert\Phi_{\theta}\rangle\propto a\vert\Psi_{\theta}\rangle$,
for same amplitude and squeezing parameters listed in Table \ref{tab:2}
and Table \ref{tab:4}. The results are shown in Fig. \ref{fig:12}
(a) and \ref{fig:13} (b) as black curves. We observe that over the
entire considered photon-loss range ( $0\le\kappa t\le0.02$), the
discrepancy between our actual fidelity $F$ and the ideal loss-channel
fidelity remains below $1\%$. This indicates that our generated cat
states retain structured phase-space distributions nearly identical
to those of the ideal single-photon-loss case, and may remain operationally
useful in lossy environments \citep{Teo_2025,hyk9-zxt2}.

\section{\label{sec:6}Conclusion }

In summary, we have proposed an OPA-based heralding framework that
generates high-fidelity CPSs and amplifies Schrödinger cat states
from $\alpha_{in}\le1$ to $\alpha_{out}\ge2$ with $F\ge0.99$, using
both catalytic ( $m=n$) and non-catalytic ( $m\neq n$) configurations.
The catalytic case preserves input parity and restores the idler state,
while non-catalytic cases enable parity-flipping amplification with
higher success probabilities. The amplified output can serve as a
seed for subsequent rounds, providing a self-seeding pathway to progressively
larger cat states. Requiring only moderate OPA gain and low-order
photon detection, our protocol offers a flexible and experimentally
accessible platform for non-Gaussian state engineering, with enhanced
robustness against photon loss due to the squeezed nature of the output
states.

Looking forward, our results suggest a practical route toward fault-tolerant
continuous-variable quantum computation: the amplified squeezed cat
states can serve as building blocks for GKP state preparation and
bosonic error correction \citep{PhysRevX.8.021054,PRXQuantum.2.020101,PhysRevA.106.022431},
while the ability to generate CPSs from coherent light provides a
direct pathway to universal Gaussian-assisted quantum computation
\citep{RevModPhys.77.513,Walschaers_2023}. Importantly, the self-seeding
capability offers a scalable strategy for generating increasingly
larger cat states without requiring progressively stronger squeezing
or higher-order photon detection, potentially bridging the gap between
currently available small-amplitude cat states and those required
for practical fault-tolerant quantum information processing \citep{Renault2025,My2025circuitlevelfault}.
\begin{acknowledgments}
This study was supported by the National Natural Science Foundation
of China (No. 12365005).
\end{acknowledgments}

\bibliographystyle{apsrev4-1}
\bibliography{Reference}

\end{document}